%% file: main.tex
\documentclass[letterpaper,twocolumn,10pt]{article}
\usepackage{usenix,epsfig,endnotes}

\usepackage{booktabs}
\usepackage{enumitem}
\usepackage{amsmath, amssymb}
\usepackage{rotating}
\usepackage{todonotes}
\microtypecontext{spacing=nonfrench}

\begin{document}

\date{}

\title{\Large \bf Making Acoustic Side-Channel Attacks on Noisy Keyboards Viable with LLM-Assisted Spectrograms' ``Typo'' Correction
}

\author{
  Seyyed Ali Ayati\thanks{These authors contributed equally.} \\ 
  \textit{ali.a@tamu.edu} \\
  Texas A\&M University
  \and
  Jin Hyun Park\footnotemark[1] \\
  \textit{jinhyun.park@tamu.edu} \\
  Texas A\&M University
  \and
  Yichen Cai \\
  \textit{yichen@cs.toronto.edu} \\
  University of Toronto
  \and
  Marcus Botacin \\
  \textit{botacin@tamu.edu} \\
  Texas A\&M University
}

\maketitle


\subsection*{Abstract}
\input{Sections/abs}
\input{Sections/intro}
\input{Sections/motiv}
\input{Sections/back}
\input{Sections/meth}

\input{Sections/eval}
\input{Sections/disc}
\input{Sections/rel}

\input{Sections/conc}

\footnotesize \bibliographystyle{acm}
\bibliography{main}

\end{document}

%% file: Sections/abs.tex
The large integration of microphones into devices increases the opportunities for Acoustic Side-Channel Attacks (ASCAs), as these can be used to capture keystrokes' audio signals that might reveal sensitive information. However, the current State-Of-The-Art (SOTA) models for ASCAs, including Convolutional Neural Networks (CNNs) and hybrid models, such as CoAtNet, still exhibit limited robustness under realistic noisy conditions. Solving this problem requires either: (i) an increased model's capacity to infer contextual information from longer sequences, allowing the model to learn that an initially noisily typed word is the same as a futurely collected non-noisy word, or (ii) an approach to fix misidentified information from the contexts, as one does not type random words, but the ones that best fit the conversation context. In this paper, we demonstrate that both strategies are viable and complementary solutions for making ASCAs practical. 
We observed that no existing solution leverages advanced transformer architectures' power for these tasks and propose that: (i) Visual Transformers (VTs) are the candidate solutions for capturing long-term contextual information and (ii) transformer-powered Large Language Models (LLMs) are the candidate solutions to fix the ``typos'' (mispredictions) the model might make. Thus, we here present the first-of-its-kind approach that integrates VTs and LLMs for ASCAs.

We first show that VTs achieve SOTA performance in classifying keystrokes when compared to the previous CNN benchmark. 
Second, we demonstrate that LLMs can mitigate the impact of real-world noise. Evaluations on the natural sentences revealed that: (i) incorporating LLMs (e.g., GPT-4o) in our ASCA pipeline boosts the performance of error-correction tasks; and (ii) the comparable performance can be attained by a lightweight, fine-tuned smaller LLM (67 times smaller than GPT-4o), using Low-Rank Adaptation (LoRA). Our results and findings highlight the practical viability of our solution toward effective ASCA.

%% file: Sections/intro.tex
\section{Introduction}
\label{sec:intro}

Acoustic Side-Channel Attacks (ASCAs) exploit the acoustic emanations from electronic devices to infer sensitive information without directly accessing the targeted systems. This adversarial method represents a substantial cybersecurity risk, particularly today as microphones become ubiquitous components integrated into common consumer devices, such as smartphones, laptops, digital assistants, and conferencing equipment. Although initially perceived as a novel theoretical threat, practical acoustic attacks have emerged to compromise sensitive information in various contexts ~\cite{zhuang2009keyboard, carrara2015acoustic, compagno2017don, mehrnezhad2018stealing}.

Previous research has demonstrated the feasibility of recovering typed passwords and PINs covertly by analyzing keystroke sounds~\cite{zhuang2009keyboard, asonov2004keyboard, berger2006dictionary, anand2018keyboard, harrison2023practical}, reconstructing confidentially printed text from dot-matrix printer noises in sensitive environments such as medical and banking facilities~\cite{backes2010acoustic}, and extracting cryptographic information or CPU operation details from subtle hardware acoustic signals~\cite{shamir2004acoustic, ninan2024second}. Among these, keyboard-focused acoustic attacks have been proven to be especially threatening due to their widespread applicability and ability to compromise passwords and confidential communications effectively~\cite{zhuang2009keyboard, asonov2004keyboard, harrison2023practical, halevi2012closer}.

Early ASCA methodologies primarily depended on straightforward statistical analyses and signal-processing techniques, leveraging acoustic features like cepstrum coefficients and inter-keystroke timing patterns for keystroke recovery~\cite{zhuang2009keyboard}. Despite enhancements through cross-correlation analysis and time-based localization techniques that improved keystroke identification~\cite{berger2006dictionary, de2019differential}, these approaches exhibited significant limitations when faced with realistic scenarios involving ambient noise. Machine-Learning (ML) methods such as Hidden Markov Models (HMMs) and Support Vector Machines (SVMs) subsequently improved accuracy by modeling sequential structures and key feature distinctions~\cite{backes2010acoustic,wang2016accurate}.

However, these early systems remained vulnerable to real environmental conditions, where significant amounts of ambient noise and recording artifacts could drastically degrade results—potentially causing incorrect keystroke identifications. Previous research has shown accuracy reductions of 30–50\% in noisy environments~\cite{halevi2012closer, harrison2023practical}. Our experiments confirm this, with accuracy dropping over 50\% under high ambient noise. Such inaccuracies are critical, particularly when attackers aim to recover sensitive information such as passwords, where even minor incorrect inferences make full password reconstruction near-impossible. Recent Deep-Learning (DL) approaches, notably Convolutional Neural Networks (CNNs) and architectures like CoAtNet~\cite{dai2021coatnet}, have further advanced acoustic classification accuracy by enabling effective hierarchies of acoustic feature extraction~\cite{harrison2023practical}. 

Despite these latest improvements, the fundamental challenge--vulnerability to noisy environmental conditions--remains largely unaddressed in the existing literature. Moreover, incorrect keystroke predictions under such conditions render traditional ASCA pipelines significantly less practical and reliable. Therefore, solutions for the noisy problem are still warranted. Our key observation and insight about it is that solving this problem requires either: \textbf{(i)} an increased model's capacity of inferring contextual information from longer sequences, allowing the model to learn that an initially noisily typed word is the same as a futurely collected non-noisy word, making the model learn a noisy representation for the characters directly from noisy data; or \textbf{(ii)} an approach to fix misidentified information from the contexts, exploiting the fact that one does not type random words, but the ones that best fit the conversation context. In this case, we would train the model with non-noisy data and allow the model to make a wrong prediction from the noisy spectrogram, a spectrogram ``typo'', but we would fix the prediction based on the context. However, technical solutions to deploy these strategies were not available in the literature until recently.

At the same time we identified the above limitations and needs, transformer architectures have emerged as a powerful class of neural network solutions capable of capturing complex long-range correlations and contextual relationships within data. Originally developed for sequence modeling in the language domain, transformers have demonstrated remarkable performance across multiple modalities—including language (LLMs), vision (VTs), and more recently audio—by leveraging their ability to globally correlate data patterns. In particular, vision transformers (e.g., ViT~\cite{dosovitskiy2020image} and Swin~\cite{liu2021swin}) successfully model spatial correlations in images and spectrogram representations, while large language models (e.g., GPT-3~\cite{brown2020language} and GPT-4~\cite{achiam2023gpt}) excel at contextual inference, error detection, and text correction tasks. These advantages naturally position transformers as candidate solutions for deploying our above-mentioned strategies. Despite these advantages, to the best of our knowledge, no research has yet leveraged transformer architectures—either vision or language variants—to handle noisy acoustic keystroke data. This position our work as the first-of-its-kind in making ASCAs practical.

Considering the above, we postulate three key scientific hypothesis: \textbf{(1)} The global modeling capability of visual transformers, particularly their unparalleled skill in capturing long-range contextual relationships, might effectively mitigate errors introduced by acoustic noise and improve ASCA reliability, making them the most suitable candidate for solving the problem via the first above-pointed strategy \textit{(i)}; and, similarly \textbf{(2)} Large Language Models (LLMs) have significant error correction (e.g., rewritting, passphrase, translation, and so on) capabilities, which makes them the most suitable solutions to solve the problem via the second above-mentioned strategy \textit{(ii)}; finally, \textbf{(3)} These solutions can complement each other and be used in tandem to solve the problem at a higher scale.

To test our hypothesis in practice, we developed a novel transformer-driven framework for enhancing acoustic side-channel attacks on keyboards and conducted experiments with it on reference datasets. More specifically, we leveraged advanced transformer-based architectures, employing both VTs (for spectrogram image classification) and LLMs (for contextual error correction), to substantially improve accuracy and robustness in noisy real-world settings. Our framework will be made available as open-source. \footnote{\url{https://github.com/seyyedaliayati/EchoCrypt}}

Our experimental evaluation was conducted using the Phone (keystrokes recorded via a smartphone microphone) and Zoom (keystrokes captured through Zoom audio call) datasets ~\cite{harrison2023practical}, considering two realistic settings. To simulate noise conditions, we introduced low, medium, and high noise levels. We compared GPT-4o~\cite{hurst2024gpt} with Llama-3.2-3B (fine-tuned) and tested Llama-3.2-1B, 3B, and 8B (non-fine-tuned)~\cite{touvron2023llama} to assess their effectiveness in mitigating noise-related errors. 

In sum, the contributions in this work are the following:

\begin{itemize}
    \item We design, implement, and evaluate VT-based frameworks for ASACs based on the Swin transformer and the CoAtNet model and demonstrate how their hypothesized capabilities of identifying long-range patterns enable them to establish the new SOTA performance in the reference ASAC benchmark.
    \item We design, implement, and evaluate LLM-based frameworks for ASACs based on GPT-4 and demonstrate that their hypothesized correction abilities allow extending the previous VT operation to noisy scenarios.
    \item We demonstrate that fine-tuned lightweight LLMs (via Low-Rank Adaptation, LoRA) can achieve typo-correction performance comparable to very large LLMs like GPT-4o, while having substantially fewer parameters—making these solutions practical even for resource-constrained attack scenarios.
\end{itemize}

In sum, the main findings of this work are the following:

\begin{enumerate}
   \item Our proposed VT framework achieved a new SOTA performance on the reference dataset, with an increase in absolute precision of 5.0\% on the Phone dataset and 5.9\% on the Zoom dataset compared to the previous CNN baselines.
   \item Our proposed LLM framework, incorporating GPT-4o to handle noisy cases, when applied to the \textit{EnglishTense} dataset (1K sentences), significantly boosts BLEU scores from approximately 0.07 to 0.90 under intermediate noise conditions, transforming an impractical classification case into a case with almost perfect key recovery.
   \item Our proposed fine-tuned LLM (Llama-3.2-3B), via Low-Rank Adaptation (LoRA), achieves accuracy scores closely comparable to GPT-4o (98-99\% of it) while being 67x smaller, which makes ASACs not only practical in noisy scenarios but also brings ASACs to one's pocket.
\end{enumerate}

The remainder of this paper is structured as follows. Section~\ref{sec:motiv} presents a clear example of our solution's goals and potential. Section~\ref{sec:background} presents background information to support our developments towards making ASCAs practical. Section \ref{sec:method} details our methodology, including our proposed vision transformer and LLM approaches, data preprocessing strategies, noise simulation methods, fine-tuning procedures, and experimental setups. Section \ref{sec:evaluation} presents our evaluation results, highlighting comparative analysis between previous state-of-the-art methods and our approach across two datasets and multiple metrics, and demonstrating the effectiveness of our proposed transformer architectures under noisy conditions. Section \ref{sec:discussion} provides an in-depth discussion of our results, practical implications, and limitations, outlining directions for future research. Section~\ref{sec:related} positions our contributions among related work, highlighting previous methodologies in acoustic keyboard side-channel classification, the evolution of classification techniques, and the emergence of transformer-based methods. Finally, Section \ref{sec:conclusion} concludes the paper by summarizing our key contributions and findings.

%% file: Sections/motiv.tex
\section{Motivating Example}
\label{sec:motiv}
\begin{figure}[htbp]
    \centering
    \includegraphics[width=\columnwidth]{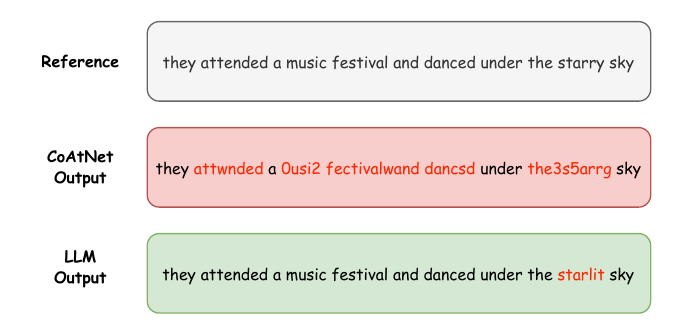}
    \caption{Example of Error Detection and Correction Using LLMs: The initial text sequence (top) represents the ideal output. The noisy prediction (middle) introduces typographical and semantic errors due to environmental noise or model inaccuracies. The corrected output (bottom) demonstrates how LLMs refine the sequence using contextual understanding, substituting errors (e.g., \textit{attwnded} to \textit{attended}).}
    \label{fig:example}
\end{figure}

To understand the goal and the potential of our solution, consider the case of the ASCAs attack shown in Figure~\ref{fig:example}.

The first line displayed in Figure~\ref{fig:example} represents the text typed by the victim. Most works on the literature consider non-noisy scenarios; thus, this same text is recovered, which makes them claim high recovery rates. In noisy scenarios, however, the recovered text will be similar to the one displayed in the second line of Figure~\ref{fig:example}, which explains why most works fail in recovering the text in these scenarios. Ideally, we would like to have a correction mechanism that allows this text to be mapped back to the original text as much as possible, as shown in the last line. 

\begin{figure}[htbp]
    \centering

    \includegraphics[width=0.50\columnwidth]{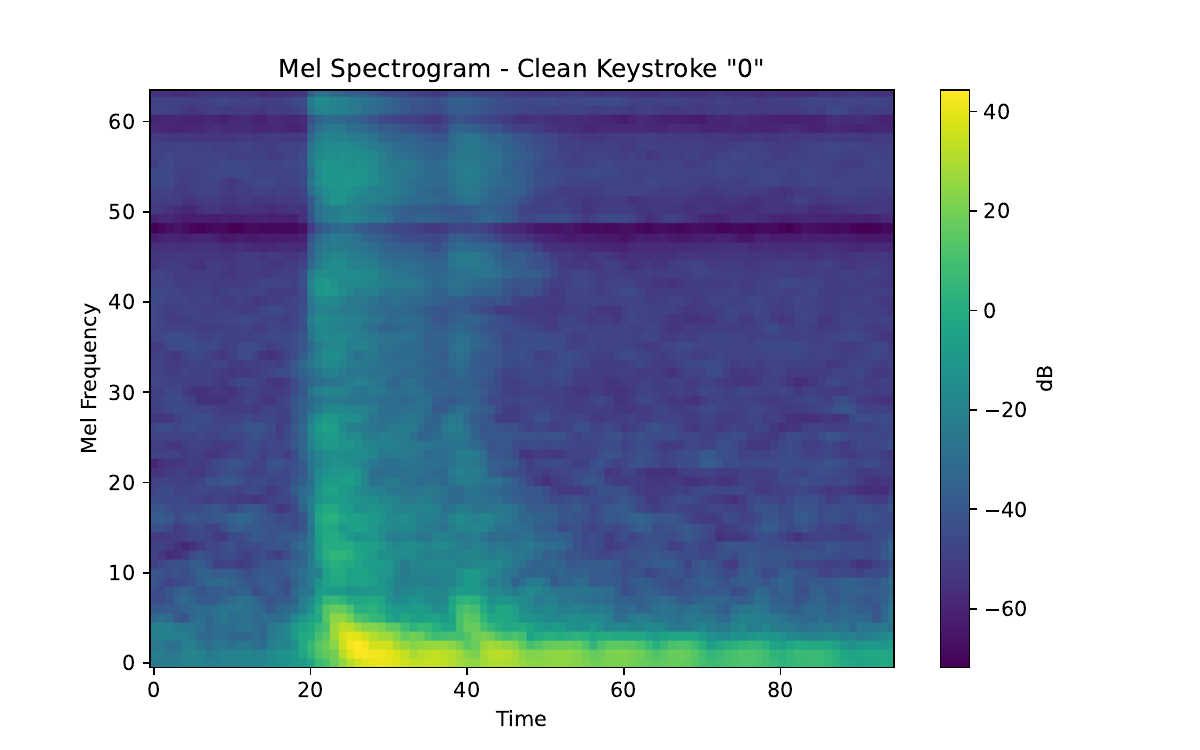}\hfill
    \includegraphics[width=0.50\columnwidth]{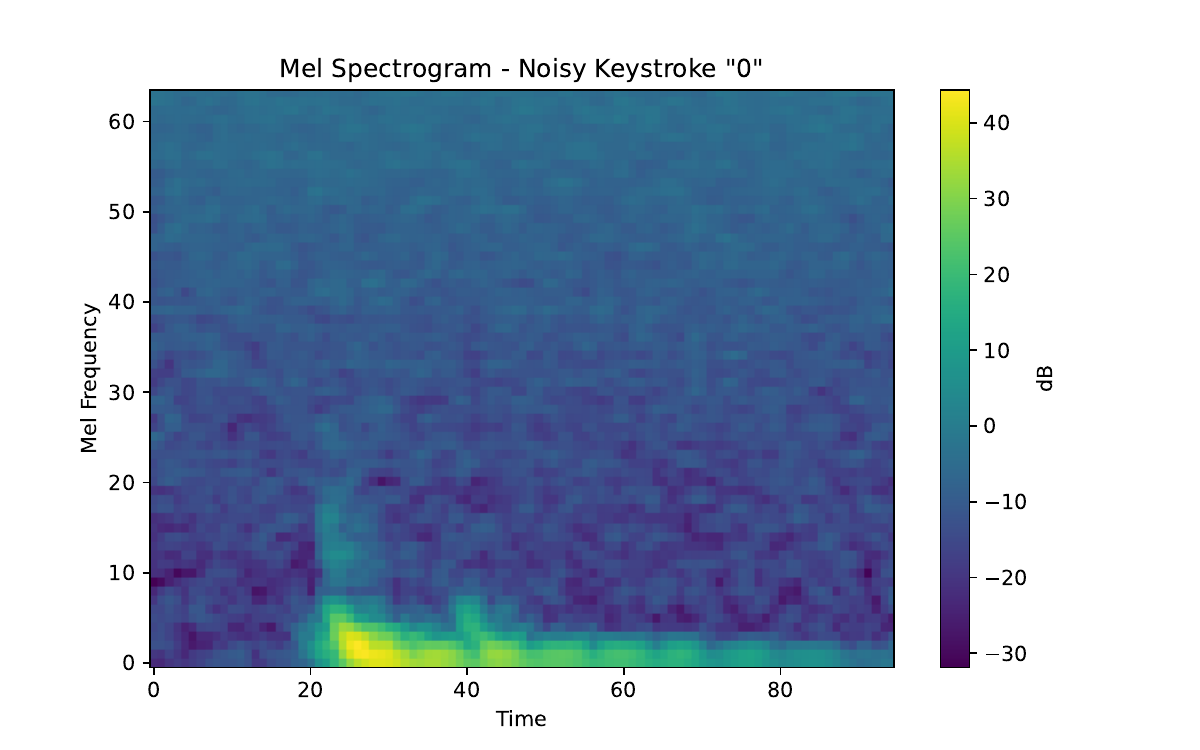}

    \caption{Comparison of Mel spectrograms for the keystroke corresponding to digit ``0'' from the phone dataset in a clean (left) and noisy (right) scenario. The noisy spectrogram exhibits additional artifacts that distort and attenuate the keystroke's spectral features, reducing the signal energy and making classification significantly more challenging.}
    \label{fig:compare}
\end{figure}

While textual examples illustrate the impact of noise, it is also insightful to visualize the differences at the signal level.Figure~\ref{fig:compare} presents two Mel spectrograms of the same keystroke—specifically, the digit ``0'' from the phone dataset—one in a clean setting and the other in a noisy setting. The spectrogram on the left corresponds to the clean scenario, where the keystroke exhibits distinct frequency characteristics. On the right, background noise distorts these features, making classification significantly harder. This visual evidence supports our argument that a robust classification and correction mechanism is necessary to achieve reliable text recovery in real-world settings.

These presented examples were obtained by applying the proposed tool to the reference benchmark, which positions our solution closer to the ideal scenario than the previous related works. This demonstrates the effectiveness of our approach in handling noisy conditions, ensuring higher accuracy in text recovery. The following sections describe how this result can be achieved.

%% file: Sections/back.tex
\section{Background}
\label{sec:background}

In ASCA, keystroke classification relies on transforming raw audio signals into representations like mel-spectrograms, and training models to identify the acoustic patterns within them. Recent developments in deep learning offer architectures specifically tailored for image data, including spectrograms~\cite{yang2023slnet, esmaeilpour2020sound, taheritajar2024survey}. Several key architectures are discussed in this paper and introduced in the following.

\textbf{CoAtNet} (Convolutional Attention Network)~\cite{dai2021coatnet} is a hybrid architecture that integrates Convolutional Neural Networks (CNNs) with self-attention mechanisms. It leverages the strength of both architectures. The convolutional layers are suitable for extracting local spatial features, e.g., edges and textures in the mel-spectrogram for keystroke audios. The self-attention mechanism captures long-range dependencies across the entire input. It focuses on the temporal relationships in the keystrokes. CoAtNet's capability of analyzing local and long-term features is promising for the ASCA task. Its staged design by stacking convolutional blocks before transformer layers enables efficient hierarchical feature extraction.

\textbf{Swin Transformer} enhances Vision Transformers (VTs) by introducing a hierarchical, shifted-window approach to self-attention~\cite{liu2021swin}. Unlike standard VTs that treat images as sequences of patches and apply global self-attention, Swin partitions images into non-overlapping patches and computes self-attention within local windows, significantly reducing computational complexity. It periodically shifts these windows, enabling partial overlaps that facilitate global context aggregation. Furthermore, Swin Transformer adjusts patch and window sizes at different stages, allowing it to efficiently capture varying levels of detail and improving both scalability and performance in vision tasks.

\textbf{Audio models,} particularly those designed for speech recognition, such as OpenAI Whisper~\cite{radford2023robust} and Mozilla DeepSpeech~\cite{hannun2014deep}, are primarily built for tasks like speech-to-text or text-to-speech conversion. While these models excel at processing linguistic content, they are not optimized for detecting short-duration, transient events like keystrokes. As a result, their pre-trained models cannot be directly applied to the task of ASCA on keyboards unless a large dataset containing long sequences of keystrokes is available.

\textbf{Architecturally, these models share similarities with the method proposed in this work}, as they also transform audio signals into mel-spectrograms and analyze the resulting visual representations. However, due to the need to process a significantly larger vocabulary, these models are inherently heavyweight. Consequently, this work does not utilize general-purpose audio-to-text models.

%% file: Sections/meth.tex
\section{Methodology}
\label{sec:method}

We here describe the methodological approaches we designed for making ASCAs viable under noisy conditions. We initially present an overview of our methodology for handling noisy data in the proposed pipeline. Later, we discuss the required steps to implement our strategy. The steps are three. First, we describe how we built a SOTA baseline for the ASCAs task, benefiting from the long-range correlation capabilities of VTs. Second, we show how to apply LLMs to fix the spectrogram's ``typos'' with their interpretation abilities. Finally, we show how to fine-tune a model to bring it to the user's pockets.

\subsection{Threat Model}

This work considers two attack scenarios. First, \textbf{Phone Recording}, in which an attacker covertly places a smartphone or microphone-equipped device near a target’s keyboard to record keystroke sounds. Modern smartphone microphones effectively capture detailed acoustic signals from short distances. The attack assumes sensitive typing activity, such as passwords or confidential messages, occurring unnoticed by the victim. Such scenarios are realistic in public spaces, offices, or co-working environments where a hidden device can passively collect keystroke audio.

Second, \textbf{Zoom Recording}, executed remotely by capturing keystroke sounds transmitted through conferencing applications such as Zoom, MS Teams, or Google Meet. Unlike phone recordings, this method doesn't require physical proximity. Laptop microphones frequently transmit keystroke sounds despite built-in noise suppression, making residual acoustic information available. Given widespread remote work and virtual meetings, attackers monitoring audio streams can reconstruct sensitive inputs using machine learning. Studies indicate keystroke sounds persist even amid compression and bandwidth constraints, enhancing attack feasibility~\cite{compagno2017don}.

The adversary is assumed to have computational resources and expertise in deep learning-based ASCA methodologies. The attack involves capturing keystroke audio, generating Mel spectrograms, and classifying keystrokes using a transformer-based model (VTs) capable of modeling complex dependencies. To address potential errors due to noise, an LLM-based correction mechanism leverages contextual understanding to refine predictions, significantly improving keystroke recovery accuracy.

The success of these attacks depends on microphone sensitivity, typing speed, keyboard type, and environmental noise. Under controlled conditions, our VT-based classification achieves over 96\% accuracy, surpassing CNN-based techniques. In noisy conditions, integrating LLM-based error correction raises text recovery accuracy from approximately 50\% to over 90\%. These findings highlight vulnerabilities in keystroke-based systems and raise broader concerns for password authentication and digital communication security.

\subsection{Overview: Handling Noise}
\begin{figure*}[htbp]
    \centering
    \includegraphics[width=\textwidth]{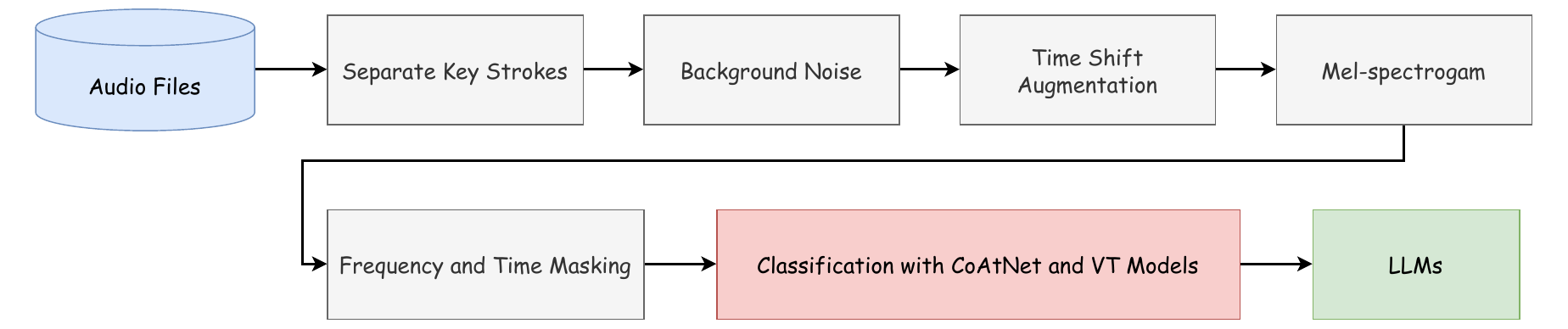}
    \caption{Audio pre-processing pipeline for classification and LLM-based typo correction.}
    \label{fig:data_processing}
\end{figure*}

A key innovation in our design is the introduction of controlled noise in the audio signals, combined with additional random time-shifting of audio signals and random time and frequency masking applied to the mel-spectrogram representations. Specifically, we employ Gaussian noise at three levels—low, medium, and high—to analyze model robustness under challenging conditions. The noise is added directly to the waveform signal before the features are extracted. This makes sure that the degradation is realistic and sounds like background noise, such as chatter or hum.

The rationale behind introducing random time shifts is to account for the natural variations in keystroke timing that occur due to human typing inconsistencies and microphone delays. This step prevents models from overfitting to precise timing patterns, improving their generalization to real-world conditions. Similarly, the frequency and time masking that was used on the mel-spectrograms simulates the loss of information that can happen during real-life recordings, where spectral components may be partially blocked by the placement of the microphone or other sounds in the background.

In our experiments, noise augmentation was applied to the entire dataset—both during training and inference—to comprehensively evaluate model robustness. This ensures that the model encounters realistic noise conditions throughout the pipeline, rather than relying solely on clean training data.

Figure~\ref{fig:data_processing} provides an overview of the preprocessing pipeline. It shows how raw audio is turned into mel-spectrograms, which are then enhanced by steps such as adding noise, shifting time, and masking frequencies before classification. The figure also highlights the integration of LLMs for correcting noisy keystroke predictions, ensuring robust and contextually accurate outputs.

\subsection{Building a SOTA Baseline model}

Prior to addressing noise scenarios, we first developed a new SOTA model to be used as a baseline for performance measurements. Our concern is not to attribute to noise effects the impact actually caused by poor model performance. Therefore, we ensured to first build a reliable baseline that allowed us to isolate variables and further focus exclusively on noisy cases.

To build the new SOTA model, we utilized CoAtNet alongside five state-of-the-art vision transformer models: ViT~\cite{dosovitskiy2020image}, Swin~\cite{liu2021swin}, DeiT~\cite{touvron2021training}, CLIP~\cite{radford2021learning}, and BEiT~\cite{bao2021beit}. We selected CoAtNet for its hybrid architecture, combining convolutional and transformer-based processing. This architecture is ideal for mel-spectrogram classification, as it captures both local features and long-range dependencies.

Each additional VT model was included to ensure comprehensive evaluation, with each offering distinct advantages. ViT serves as a pure transformer-based benchmark, while Swin introduces efficient hierarchical feature extraction via shifted windows. DeiT is optimized for data efficiency, suited for limited-sample datasets. CLIP was tested for potential performance gains from its vision component, despite its multimodal nature. BEiT's masked image modeling pre-training allowed us to assess the benefits of strong visual pretraining in keystroke classification. This diverse selection enabled systematic evaluation and ensured our final choice was driven by empirical performance rather than arbitrary selection.

We used pre-trained VT models and fine-tuned them with our data. For the CoAtNet model, since the previous study's ~\cite{harrison2023practical} CoAtNet (it will be refered to it as B-CoAtNet for baseline CoAtNet) model does not specify the number of blocks or channel sizes, we configured the number of parameters to align with the smallest reported parameter size in the CoAtNet's original paper ~\cite{dai2021coatnet}. Notably, unlike the other selected VTs, CLIP is a multimodal model, meaning it incorporates both vision and language embeddings for downstream tasks. In our experiments, we exclusively utilized the vision component of the CLIP model, augmenting it with a fully connected layer before the softmax layer. 

\subsection{Finding and Fixing Errors with LLMs}
Detecting and correcting errors in noisy textual predictions is challenging due to the unpredictable nature of acoustic distortions. Unlike traditional spell-checking tasks, where errors follow common typing patterns, ASCA errors involve random character substitutions, deletions, or insertions caused by waveform perturbations. Thus, an LLM must infer missing or altered characters based on broader linguistic context, reconstructing meaningful sequences even without clear word boundaries (e.g., missing spaces). Our approach leverages few-shot prompting to adapt LLMs effectively to these distortions, ensuring robust correction across varying noise levels and keystroke misclassification patterns.

Detecting and correcting errors in noisy textual predictions is a critical step in enhancing the robustness of ASCA systems. This section details the methodology employed to address errors arising from noisy datasets by leveraging LLMs through few-shot prompting techniques. The process involves both datasets, noise environments, and evaluation metrics, providing a comprehensive analysis of error correction effectiveness. We did not consider a noise-free environment, as it is highly unlikely to encounter zero noise in real-world scenarios.

The selected sentences were mapped to acoustic data under different noise environments to simulate real-world scenarios, as shown in Table \ref{tab:noise_factors}. Noise factors were categorized as Low, Medium, and High. For each syllable or digit in a sentence, the corresponding sound wave was adjusted with noise using different noise factors. See Eq.~\ref{eq:add_noise} to see how noises are applied.
\begin{equation}
    S_{\text{noisy}} = S + \eta \cdot \mathcal{N}(0, 1)
\label{eq:add_noise}
\end{equation}
where $S$ indicates acoustic signal (i.e. sound wave) and $\eta$ is a noise factor.
We carefully selected noise factor values to ensure that the resulting accuracies for Low, Medium, and High noise levels were approximately 95\%, 85\%, and 70\%, respectively. The accuracy was calculated between the true sentence and sentence with errors. See Eq.~\ref{eq:accuracy} for more details. 
\par
\begin{equation}
    \text{Accuracy} = \frac{2 \cdot |M|}{|S_1| + |S_2|}
\label{eq:accuracy}
\end{equation}
\text{where:}
\begin{itemize}
    \item \( |M| \): The total number of matching characters between the two strings \( S_1 \) (the true sentence) and \( S_2 \) (the predicted sentence). Matches are determined by optimizing the alignment of characters, not restricted to contiguous sequences.
    \item \( |S_1| \): The length (number of characters) of the true sentence \( S_1 \).
    \item \( |S_2| \): The length (number of characters) of the predicted sentence \( S_2 \).
\end{itemize}
\begin{figure}[htbp]
    \centering
    \includegraphics[width=\columnwidth]{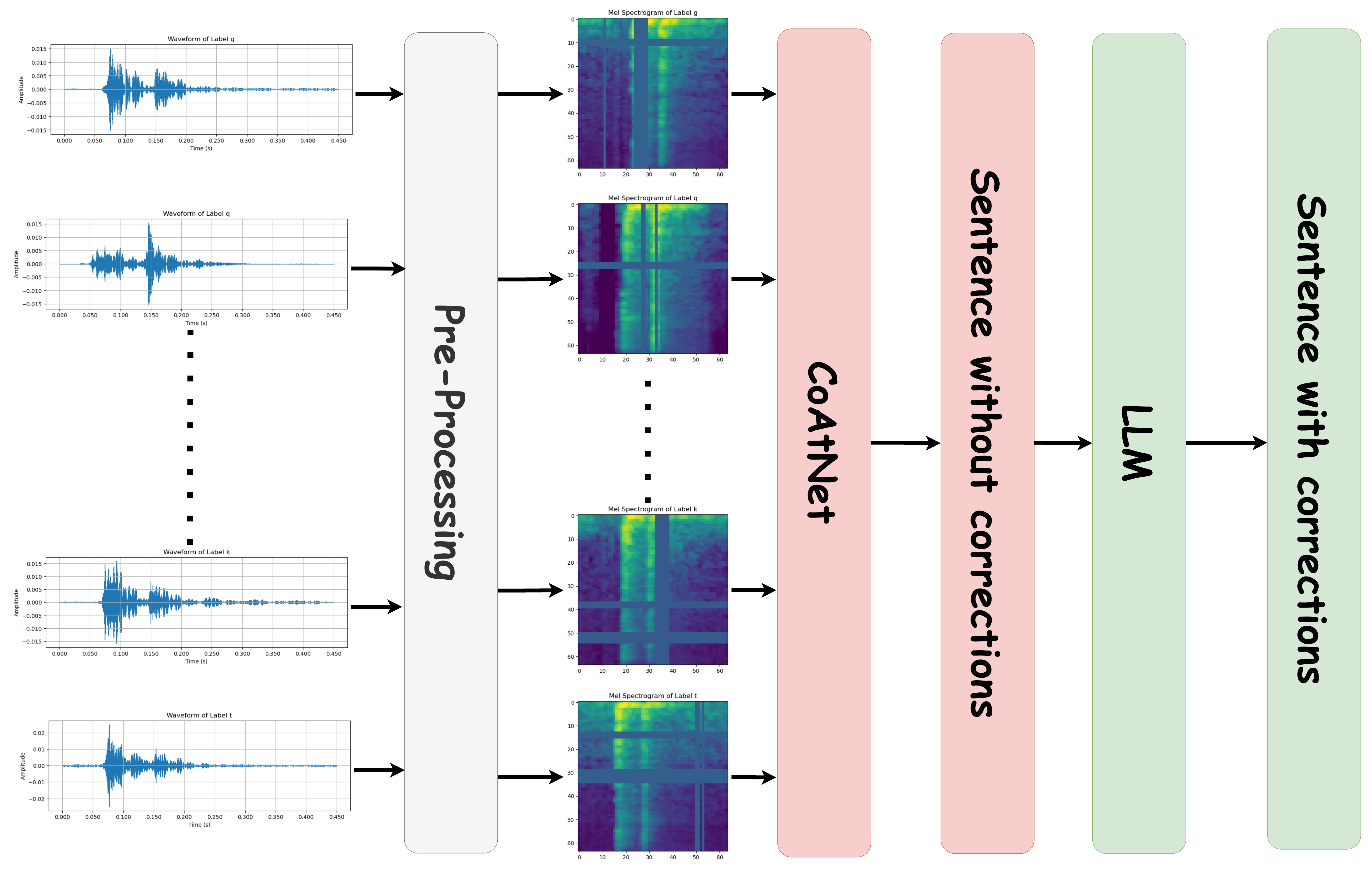}
    \caption{Pipeline for Detecting and Correcting Errors in Keystroke Predictions Using LLMs: From noisy audio waveforms to mel-spectrogram processing, keystroke classification via CoAtNet, and error detection/correction with LLMs, resulting in an accurate and more probable textual output.}
    \label{fig:pipeline}
\end{figure}
\par
The noisy audio waveforms were processed through the pipeline illustrated in Fig.~\ref{fig:pipeline}. After generating noisy textual predictions, typographical and semantic errors were corrected using LLMs. Few-shot prompting techniques, as outlined in Table~\ref{tab:prompt_structure}, were employed. The LLM was prompted with examples of sentences containing typos and their corrected counterparts. 
We utilized Llama-3.2-1B, Llama-3.2-3B, Llama-3.1-8B, and GPT-4o for LLM models.
The system prompt established the role of the LLM as an expert in correcting typos, while the user prompt included examples and the target sentence. 
\par 
\begin{table*}[htbp]
    \centering
    \begin{tabular}{p{1cm}|p{15cm}}
        \toprule
        \textbf{Role} & \textbf{Content} \\ \midrule
        \textit{System} & You are an expert in correcting typos in sentences. \\ 
        \midrule
        \textit{User} & Here are pairs of sentences with typos; learn from them: \\
             & \\
             & sentence: \{$S^{1}_{\text{pred}}$\} \\
             & corrected: \{$S^{1}_{\text{true}}$\} \\
             & \\
             & sentence: \{$S^{2}_{\text{pred}}$\} \\
             & corrected: \{$S^{2}_{\text{true}}$\} \\
             & \\
             & Now, please correct these sentences and output only the corrected version with no additional text: \{$S_{\text{pred}}$\} \\ 
        \bottomrule
    \end{tabular}
    \vspace{0.2em}
    \caption{Few-shot prompting structure for generating messages to correct typos using LLMs. The \textit{System} role sets the task context, and the \textit{User} role provides pairs of sentences (1) before correction $S_{\text{pred}}$ (sentence with errors) and (2) after correction $S_{\text{true}}$ (sentence without errors)}
    \label{tab:prompt_structure}
\end{table*}

\subsection{Detecting and Correcting Errors with Fine-tuned LLMs}
We hypothesized that GPT-4o could outperform smaller models (i.e., Llama family) in accurately correcting typos and mistakes in sentences. This is because models with a high number of parameters tend to perform better than smaller models. However, a significant drawback is its high inference time, attributed to its substantial model size ($\sim$200 billion parameters). To circumvent this issue, we propose a method to create a much smaller model specifically designed for typo correction task. We used Low-Rank Adaptation fine-tuning methods ~\cite{dettmers2023qloraefficientfinetuningquantized} to achieve the objective on the smaller model.
That is, instead of fine-tuning the full set of parameters in the model, a set of low-rank update matrices is introduced on top of the pre-trained weights to train the LLM for the error correction task. This allows task-specific learning without modifying the original weights, further reducing the computational cost and memory footprint.
\par
We took a pre-trained Llama-3.2-3B ($\sim$3 billion parameters) model and chose the more efficient QLoRA method ~\cite{unsloth} for the fine-tuning task. The smaller low-rank update matrices have a dimension of $d \times c$ where $d$ is the dimension of the original weight matrices, and $c$ is chosen arbitrarily but satisfying $d$ >{}> $c$.
Training is carried out on specific training tasks where the predicted sentences from the audio are the input, and the correct output is the actual sentences being typed. To enhance the robustness of the model against noise, we progressively increase the noise level in the training data. Following Table \ref{tab:noise_factors}, the training for each dataset starts with a low noise factor and goes up to high noise intensity, one epoch for each noise level. AdamW optimizer, with a learning rate of 0.0002, is used for fine-tuning over three epochs: the first with low, the second with medium, and the third with high noise sentences.
Since we train the model on the error correction task specifically, zero-shot prompting is used. The gradient updates were only applied to the LoRA weights, while the original Llama weights were frozen. The updates can be expressed in Eq.~\ref{eq:lora_update}.
\begin{equation}
    W = W_0 + \Delta W = W_0 + AB
\label{eq:lora_update}
\end{equation}
where $W$ is the updated weight after fine-tuning, $W_0$ is the original pre-trained weight, and $\Delta W$ is the update being applied. Traditional fine-tuning directly operates on $\Delta W$, where $W, W_0, \Delta W \in \mathbb{R}^{d\times d}$ for large $d$. In contrast to traditional fine-tuning methods, we used QLoRA to operate on weighs $A, B$ where $A \in \mathbb{R}^{d\times c}, B \in \mathbb{R}^{c\times d}$ for a much smaller $c$. Instead of computing $\Delta W$ directly, we calculate $\Delta W = AB$ and only make updates on small matrices $A$ and $B$.
\par
Concluding the methodology section, our comprehensive approach demonstrates the robustness of the (fine-tuned) LLM-based error detection and correction framework across diverse noise environments and datasets. 
The results, presented in the subsequent section, demonstrate the performance improvements achieved by integrating LLMs with the ASCA pipeline, offering insights into their potential for real-world applications.

%% file: Sections/eval.tex
\section{Evaluation}
\label{sec:evaluation}

This section is organized into four subsections: (1) Data, (2) CoAtNets vs VTs, (3) Mitigating errors with LLMs, and (4) Mitigating errors with fine-tuned LLMs.

\subsection{Dataset}

The original keystroke dataset consists of two datasets, which are from the noise-free Phone and Zoom recordings with lacked space characters. To simulate real-world sentences, we assumed spaces existed in the dataset with the same prediction accuracy as other letters and digits. To evaluate error detection and correction, the EnglishTense dataset~\cite{ayman2024englishtense} was used, which contains a rich corpus of categorized English sentences. From this dataset, 1,000 sentences were randomly selected: 500 containing digits and 500 without digits, ensuring a balanced variety of sentence structures and types.
\par
\begin{table}[htbp]
\centering
\resizebox{\columnwidth}{!}{%
\begin{tabular}{l|ccc}
    \toprule
    \textbf{Dataset name / Noise Factor ($\eta$)} & \textbf{Low} & \textbf{Medium} & \textbf{High} \\ \midrule
    Phone            & 0.012            & 0.024             & 0.06             \\ 
    Zoom             & 0.1            & 0.5               & 1.0             \\ 
    \bottomrule
\end{tabular}%
}
\vspace{0.2em}
\caption{Different noise intensity levels across Phone and Zoom datasets.}
\label{tab:noise_factors}
\end{table}
\par

We utilized the same datasets as in ~\cite{harrison2023practical}. The data were collected from two sources: Phone and Zoom ~\cite{datasetgithublink}, which contain recordings of 36 distinct keystrokes (letters a–z and digits 0–9), with each keystroke recorded 25 times using either setup. 
In the Phone setup, the sound of a victim’s laptop keystrokes was recorded using an iPhone (representing the attacker). In the Zoom setup, the victim’s laptop keystroke sounds were captured via Zoom's built-in recording function, with the audio obtained from the victim’s microphone.
We selected this dataset over other available options for several reasons: (1) It is a recent dataset (2023) specifically designed for acoustic side-channel attacks; (2) it employs recordings from a widely available, modern laptop, the MacBook Pro 16-inch (2021), whose keyboard design remains prevalent in current (2025) MacBook models; and (3) the Zoom recording setup offers a particularly realistic scenario, given the widespread use of Zoom today. 
\par 
However, a significant limitation of this dataset is that all recordings were made in a noise-free environment. Consequently, it does not fully reflect real-world scenarios, and a standalone classification model is likely to struggle when confronted with even minimal background noise, a challenge we address using an LLM-based correction mechanism. Additionally, the dataset does not include \textit{space}, \textit{backspace}, and \textit{enter} keystrokes, which are essential components of typical keyboard usage. We discuss these issues in more detail in Section~\ref{sec:discussion}.
\par 
Since each keystroke was recorded 25 times consecutively, every audio file was segmented into 25 individual \textit{.wav} files. To achieve this, we first applied a Fast Fourier Transform (FFT) to each recording and summed the frequency coefficients to compute the energy. This energy measure allowed us to identify 25 distinct peaks per keystroke, thereby dividing the audio into 25 separate files per key, resulting in a total of 900 samples. We followed the step shown in Fig.~\ref{fig:data_processing} to generate mel-spectrogram images. 
\par 
The input image dimensions vary between models: CoAtNet utilizes 64$\times$64 images, while VT models require 224$\times$224 images. We employed the Adam~\cite{kingma2014adam} optimizer for CoAtNet and AdamW~\cite{loshchilov2017decoupled} for VT models. Table~\ref{tab:keystroke_hyperparameters} provides detailed hyperparameters used throughout our experiments, from image generation to testing. To compare the performance of CoAtNet and VT models, we applied two distinct data transformation methods to the VT inputs. These methodologies enable a comprehensive evaluation under different preprocessing conditions:
\begin{enumerate}[label=(\arabic*)]
    \item \textbf{Resizing Transformation:} Input images initially sized at 64$\times$64 were resized using bilinear interpolation to 224$\times$224 to meet the VT models' requirements. 
    \item \textbf{Direct Transformation:} By modifying the hyperparameters of the mel-spectrogram generation process, input images were directly produced at a resolution of 224$\times$224, thereby eliminating the need for resizing. 
\end{enumerate}
\begin{table}[htbp]
    \centering
    \resizebox{\columnwidth}{!}{%
    \begin{tabular}{l|cc}
        \toprule
         & \textbf{Phone Value} & \textbf{Zoom Value} \\ \midrule
        Epochs      & 1100 & 1100 \\
        Batch Size  & 16 & 16 \\
        Loss Type   & Cross Entropy & Cross Entropy \\
        Optimizer   & Adam$^*$ / AdamW$^\dagger$ & Adam$^*$ / AdamW$^\dagger$ \\
        Max Learning Rate        & 5e-4$^*$ / 5e-5$^\dagger$ & 5e-4$^*$ / 5e-5$^\dagger$\\
        Annealing Schedule       & Linear & Linear \\
        Timeshift Percentage     & 0.3 & 0.4 \\
        Max Mask Percentage      & 0.1 $^*$ / 0.03$^\dagger$ & 0.1 $^*$ / 0.03$^\dagger$\\
        Number of Masks Per Axis & 2 & 2 \\
        Mel Bands       & 64$^*$ / 224$^\dagger$ & 64$^*$ / 224$^\dagger$\\
        FFT Window Size & 1024 & 1024 \\
        Hop Length      & 300$^*$ / 85$^\dagger$ & 226$^*$ / 64$^\dagger$\\
        Data Split      & Stratified & Stratified\\
        \bottomrule
    \end{tabular}%
    }
    \vspace{0.2em}
    \caption{Hyperparameters for keystroke classification experiments using Phone and Zoom recordings. $^*$ indicates values used in our CoAtNet implementation, while $^\dagger$ indicates values used in the selected VT models}
    \label{tab:keystroke_hyperparameters}
\end{table}

\subsection{Metrics}
The corrected sentences were evaluated using multiple metrics to quantify performance:
\begin{itemize}
    \item \textbf{BLEU:} Measures n-gram overlap between the corrected sentence and the ground truth.
    \item \textbf{ROUGE:} Evaluates recall-based overlap of substrings.
    \item \textbf{METEOR:} Considers synonymy and stemming in evaluating similarity.
\end{itemize}

\subsection{Research Questions (RQs)}

\begin{enumerate}
    \item How do the different Visual Transformer (VT) architectures perform in acoustic keystroke classification compared to the state-of-the-art baseline CoAtNet under ideal (noise-free) conditions?
    \item Can Large Language Models (LLMs), such as GPT-4o and Llama models, effectively correct errors in keystroke predictions resulting from realistic acoustic noise?
    \item How does fine-tuning smaller LLMs using Low-Rank Adaptation (LoRA) affect their performance in error correction compared to larger LLMs like GPT-4o in acoustic side-channel attacks?
\end{enumerate}

\subsection{RQ1: CoAtNets vs VTs}
Using the details outlined in the methodology section, we compared B-CoAtNet with our CoAtNet (we will now refer to it as O-CoAtNet for our CoAtNet) using a noise-free dataset. As shown in Table \ref{tab:coatnet}, O-CoAtNet achieves higher accuracy than B-CoAtNet in terms of both mean and standard deviation, as well as maximum accuracy.
For the Phone dataset, we observe a 1.5\% increase in mean accuracy, while for the Zoom dataset, the improvement is 3.5\%. Given that the original work reported only the best accuracy without providing a mean or standard deviation, the gap between the best accuracy of B-CoAtNet and O-CoAtNet is even more significant. We show a 5.0\% increase in the Phone dataset, while for the Zoom dataset, the improvement is 5.9\%.
Given that we utilized the lightweight CoAtNet model, we anticipate that even higher accuracy could be achieved with a heavyweight CoAtNet model, as previous work ~\cite{dai2021coatnet} reports better performance for heavyweight CoAtNet configurations under ImageNet ~\cite{5206848} and JFT-300M ~\cite{sun2017revisiting} datasets.
\begin{table}[htbp]
    \centering
    \resizebox{\columnwidth}{!}{%
    \begin{tabular}{l|cc}
        \toprule
        & \textbf{B-CoAtNet (Unknown)} & \textbf{O-CoAtNet ($\sim$24M)} \\ \midrule
        Phone: Mean and stdev & - & 96.45 $\pm$ 3.5\% \\
        Phone: Max  & 95\% & \textbf{100\%} \\
        \midrule
        Zoom: Mean and stdev & - & 96.67 $\pm$ 2.1\% \\
        Zoom: Max  & 93\% & \textbf{98.9\%} \\
        \bottomrule
    \end{tabular}%
    }
    \vspace{0.2em}
    \caption{Summary statistics for B-CoAtNet vs O-CoAtNet on the noise-free Phone and Zoom datasets. For our CoAtNet, the mean and standard deviation were calculated based on results from five different seeds. Note that B-CoAtNet only reports the best accuracy without providing standard deviations or parameter sizes, the latter of which is marked as `unknown' in the table.}
    \label{tab:coatnet}
\end{table}
\par
We also compared B-CoAtNet with selected VTs, including a reference to the number of parameters for each model. By examining the results in Tables \ref{tab:trans1}, we observe that VTs demonstrate comparable or, in some cases, superior accuracy to B-CoAtNet. 
When comparing the best accuracy between B-CoAtNet and selected VTs, the direct transformation using Swin exhibits the largest gap, with a 5.0\% difference on the Phone dataset and a 5.9\% difference on the Zoom dataset.
Considering other models, they also demonstrate accuracy comparable to B-CoAtNet. 
However, the CLIP model exhibits significantly lower performance. We suspect this is due to CLIP being originally designed for multimodal tasks, while we utilized only its vision component. Here we want to note that, we did not fine-tune hyperparameters of VTs such as the learning rate, batch size, or scheduler, which likely contributed to the weaker performance. We strongly believe that performance could be improved through hyperparameter optimization.
\begin{table*}[htbp]
    \centering
    \small
    \begin{tabular}{p{4cm}|p{2.25cm}p{2.25cm}p{2.25cm}p{2.25cm}p{2.25cm}}
        \toprule
        \textbf{Resizing Transformation} & \textbf{ViT (86M)} & \textbf{Swin (28M)} & \textbf{DeiT (86M)} & \textbf{CLIP (87M)} & \textbf{BEiT (86M)} \\ 
        \midrule
        Phone: Mean and stdev & 94.8\% $\pm$ 1.0\% & 90.0\% $\pm$ 7.6\% & 89.3\% $\pm$ 3.3\% & 76.9\% $\pm$ 3.8\% & \textbf{95.6\% $\pm$ 1.4\%} \\
        Phone: Max & 95.6\% & \textbf{98.9\%} & 92.2\% & 81.1\% & 97.8\% \\
        \midrule
        Zoom: Mean and stdev & \textbf{90.4\% $\pm$ 3.0\%*} & 85.1\% $\pm$ 3.4\% & 85.3\% $\pm$ 2.0\% & 66.2\% $\pm$ 8.9\% & 87.1\% $\pm$ 4.1\% \\
        Zoom: Max & \textbf{94.4\%} & 88.9\% & 87.8\% & 78.9\% & 92.2\% \\ 
        \bottomrule
        \noalign{\vspace{1.0em}} 
        \toprule
        \textbf{Direct Transformation} & \textbf{ViT (86M)} & \textbf{Swin (28M)} & \textbf{DeiT (86M)} & \textbf{CLIP (87M)} & \textbf{BEiT (86M)} \\ 
        \midrule
        Phone: Mean and stdev & 94.0\% $\pm$ 2.3\% & 96.7\% $\pm$ 2.5\% & 94.4\% $\pm$ 1.4\% & 84.9\% $\pm$ 6.7\% & \textbf{97.6\% $\pm$ 1.6\%*} \\
        Phone: Max & 97.8\% & \textbf{100.0\%*} & 95.6\% & 90.0\% & \textbf{100.0\%*} \\
        \midrule
        Zoom: Mean and stdev & 56.4\% $\pm$ 48.8\% & 54.7\% $\pm$ 48.4\% & 22.2\% $\pm$ 40.4\% & 2.7\% $\pm$ 0.6\% & \textbf{57.3\% $\pm$ 45.9\%} \\
        Zoom: Max & 95.6\% & \textbf{98.9\%*} & 94.4\% & 3.3\% & 96.7\% \\
        \bottomrule
    \end{tabular}
    \vspace{0.2em}
    \caption{Summary statistics for different transformer models on the Phone and Zoom dataset using resizing transformation and direct transformation. The mean and standard deviation (stdev) were calculated using results from five different seeds. Bolded numbers indicate the best performance in each row, while an asterisk (*) denotes the best for each dataset.}
    \label{tab:trans1}
\end{table*}
\par
When evaluating the number of parameters, VTs do not appear to offer a clear performance advantage over CoAtNet, as they have a much larger parameter count. This is likely due to the limited size of the dataset, which contains only 25 samples per syllable or digit. We presume that collecting a larger dataset or employing additional data augmentation techniques could improve accuracy in VTs. Overall, our results show that VTs can perform on par with or even exceed CoAtNet. 
\par
Importantly, our goal was not merely to achieve better performance using VTs, but to assess their potential as alternative models for acoustic SCA keystroke classification. We are primarily focused on enhancing the performance of the ASCA keystroke classification task, especially under realistic conditions where environmental noise is inevitable. To address this challenge, we introduce a LLM to complement the classification process. Recognizing that relying solely on refined datasets can be impractical in noisy environments, our approach aims to improve robustness and accuracy. In the following subsection, we demonstrate how LLMs can be effectively leveraged to mitigate noise-related issues, thereby addressing more realistic operational conditions.

\subsection{RQ2: Mitigating errors with LLMs}
We present the experimental results demonstrating the performance of various LLM models in correcting noisy text predictions across multiple noise levels. 
The evaluation utilized multiple metrics, including BLEU, METEOR, and ROUGE scores, providing a comprehensive analysis of the effectiveness of LLMs in improving the robustness of ASCAs. 
We evaluated the error mitigation capabilities of LLM models, specifically the Llama family and GPT-4o, by varying error rates and performance metrics. 
\par 
\begin{table*}[htbp]
    \centering
    \small
    \begin{tabular}{l|p{2cm}p{2cm}p{2cm}p{2cm}p{2cm}p{2cm}}
        \toprule
        \textbf{Metric (Noise Factor)} & \textbf{B-CoAtNet} & \textbf{Llama-3.2-1B} & \textbf{Llama-3.2-3B} & \textbf{Llama-3.1-8B} & \textbf{GPT-4o ($\sim$200B)} &\textbf{Llama-3.2-3B (Fine-tuned)} \\ \midrule
        BLEU (Low) & 0.383 $\pm$ 0.249 & 0.684 $\pm$ 0.294 & 0.863 $\pm$ 0.183 & 0.910 $\pm$ 0.159 & 0.975 $\pm$ 0.075* & \textbf{0.977 $\pm$ 0.073} \\
        BLEU (Mid) & 0.078 $\pm$ 0.091 & 0.273 $\pm$ 0.268 & 0.542 $\pm$ 0.282 & 0.681 $\pm$ 0.258 & \textbf{0.916 $\pm$ 0.120} & 0.908 $\pm$ 0.156* \\
        BLEU (High) & 0.021 $\pm$ 0.023 & 0.041 $\pm$ 0.083 & 0.118 $\pm$ 0.151 & 0.189 $\pm$ 0.196 & \textbf{0.638 $\pm$ 0.287} & 0.623 $\pm$ 0.302* \\
        \midrule
        Meteor (Low) & 0.640 $\pm$ 0.190 & 0.830 $\pm$ 0.213 & 0.941 $\pm$ 0.084 & 0.964 $\pm$ 0.070 & 0.988 $\pm$ 0.032* & \textbf{0.990 $\pm$ 0.032} \\
        Meteor (Mid) & 0.286 $\pm$ 0.162 & 0.504 $\pm$ 0.273 & 0.751 $\pm$ 0.184 & 0.846 $\pm$ 0.143 & 0.958 $\pm$ 0.057* & \textbf{0.960 $\pm$ 0.073} \\
        Meteor (High) & 0.082 $\pm$ 0.085 & 0.137 $\pm$ 0.159 & 0.344 $\pm$ 0.211 & 0.454 $\pm$ 0.219 & 0.804 $\pm$ 0.189* & \textbf{0.807 $\pm$ 0.191} \\
        \midrule
        Rouge-1 (Low) & 0.690 $\pm$ 0.157 & 0.844 $\pm$ 0.194 & 0.944 $\pm$ 0.080 & 0.964 $\pm$ 0.071 & 0.989 $\pm$ 0.032* & \textbf{0.990 $\pm$ 0.034} \\
        Rouge-1 (Mid) & 0.374 $\pm$ 0.148 & 0.552 $\pm$ 0.249 & 0.777 $\pm$ 0.159 & 0.856 $\pm$ 0.129 & \textbf{0.960 $\pm$ 0.055} & \textbf{0.960 $\pm$ 0.071} \\
        Rouge-1 (High) & 0.141 $\pm$ 0.113 & 0.196 $\pm$ 0.173 & 0.406 $\pm$ 0.196 & 0.503 $\pm$ 0.200 & \textbf{0.817 $\pm$ 0.171} & 0.815 $\pm$ 0.176* \\
        \midrule
        Rouge-2 (Low) & 0.497 $\pm$ 0.226 & 0.754 $\pm$ 0.251 & 0.899 $\pm$ 0.137 & 0.932 $\pm$ 0.121 & 0.979 $\pm$ 0.061* & \textbf{0.982 $\pm$ 0.060} \\
        Rouge-2 (Mid) & 0.131 $\pm$ 0.136 & 0.361 $\pm$ 0.276 & 0.629 $\pm$ 0.240 & 0.748 $\pm$ 0.209 & 0.925 $\pm$ 0.103* & \textbf{0.927 $\pm$ 0.123} \\
        Rouge-2 (High) & 0.017 $\pm$ 0.050 & 0.059 $\pm$ 0.122 & 0.191 $\pm$ 0.190 & 0.285 $\pm$ 0.214 & \textbf{0.703 $\pm$ 0.244} & 0.701 $\pm$ 0.254* \\
        \midrule
        Rouge-L (Low) & 0.690 $\pm$ 0.157 & 0.844 $\pm$ 0.194 & 0.943 $\pm$ 0.080 & 0.963 $\pm$ 0.073 & 0.989 $\pm$ 0.032* & \textbf{0.990 $\pm$ 0.034} \\
        Rouge-L (Mid) & 0.374 $\pm$ 0.148 & 0.550 $\pm$ 0.249 & 0.776 $\pm$ 0.160 & 0.856 $\pm$ 0.130 & \textbf{0.960 $\pm$ 0.055} & \textbf{0.960 $\pm$ 0.071} \\
        Rouge-L (High) & 0.141 $\pm$ 0.113 & 0.194 $\pm$ 0.172 & 0.400 $\pm$ 0.196 & 0.497 $\pm$ 0.202 & \textbf{0.816 $\pm$ 0.172} & 0.814 $\pm$ 0.179* \\
        \bottomrule
    \end{tabular}
    \vspace{0.2em}
    \caption{Performance using metrics -- BLEU, METEOR, ROUGE-1, ROUGE-2, and ROUGE-L -- for different models at varying noise factors on the Phone dataset. Bolded numbers indicate the best performance (mean) in each row, while an asterisk (*) shows the second-best performance (mean) within the same metric.}
    \label{tab:metrics_noise_levels_phone}
\end{table*}
\begin{table*}[htbp]
    \centering
    \small
    \begin{tabular}{l|p{2cm}p{2cm}p{2cm}p{2cm}p{2cm}p{2cm}}
        \toprule
        \textbf{Metric (Noise Factor)} & \textbf{B-CoAtNet} & \textbf{Llama-3.2-1B} & \textbf{Llama-3.2-3B} & \textbf{Llama-3.1-8B} & \textbf{GPT-4o ($\sim$200B)} & \textbf{Llama-3.2-3B (Fine-tuned)} \\ \midrule
        BLEU (Low) & 0.285 $\pm$ 0.218 & 0.657 $\pm$ 0.297 & 0.827 $\pm$ 0.211 & 0.891 $\pm$ 0.167 & \textbf{0.976 $\pm$ 0.070} & 0.955 $\pm$ 0.099* \\
        BLEU (Mid) & 0.050 $\pm$ 0.057 & 0.193 $\pm$ 0.225 & 0.385 $\pm$ 0.273 & 0.520 $\pm$ 0.288 & \textbf{0.896 $\pm$ 0.152} & 0.866 $\pm$ 0.184* \\
        BLEU (High) & 0.018 $\pm$ 0.022 & 0.025 $\pm$ 0.055 & 0.081 $\pm$ 0.126 & 0.107 $\pm$ 0.151 & \textbf{0.542 $\pm$ 0.303} & 0.489 $\pm$ 0.303* \\
        \midrule
        Meteor (Low) & 0.551 $\pm$ 0.191 & 0.811 $\pm$ 0.211 & 0.923 $\pm$ 0.101 & 0.952 $\pm$ 0.076 & \textbf{0.988 $\pm$ 0.031} & 0.979 $\pm$ 0.043* \\
        Meteor (Mid) & 0.210 $\pm$ 0.143 & 0.412 $\pm$ 0.261 & 0.646 $\pm$ 0.216 & 0.751 $\pm$ 0.186 & \textbf{0.950 $\pm$ 0.073} & 0.938 $\pm$ 0.088* \\
        Meteor (High) & 0.065 $\pm$ 0.076 & 0.085 $\pm$ 0.117 & 0.268 $\pm$ 0.191 & 0.319 $\pm$ 0.200 & \textbf{0.735 $\pm$ 0.218} & 0.710 $\pm$ 0.226* \\
        \midrule
        Rouge-1 (Low) & 0.608 $\pm$ 0.161 & 0.826 $\pm$ 0.190 & 0.927 $\pm$ 0.092 & 0.954 $\pm$ 0.070 & \textbf{0.989 $\pm$ 0.031} & 0.979 $\pm$ 0.042* \\
        Rouge-1 (Mid) & 0.291 $\pm$ 0.144 & 0.461 $\pm$ 0.241 & 0.676 $\pm$ 0.195 & 0.769 $\pm$ 0.168 & \textbf{0.952 $\pm$ 0.070} & 0.940 $\pm$ 0.085* \\
        Rouge-1 (High) & 0.111 $\pm$ 0.102 & 0.131 $\pm$ 0.137 & 0.329 $\pm$ 0.184 & 0.373 $\pm$ 0.188 & \textbf{0.750 $\pm$ 0.199} & 0.722 $\pm$ 0.208* \\
        \midrule
        Rouge-2 (Low) & 0.392 $\pm$ 0.212 & 0.726 $\pm$ 0.254 & 0.868 $\pm$ 0.159 & 0.915 $\pm$ 0.129 & \textbf{0.979 $\pm$ 0.060} & 0.962 $\pm$ 0.079* \\
        Rouge-2 (Mid) & 0.083 $\pm$ 0.103 & 0.276 $\pm$ 0.248 & 0.493 $\pm$ 0.252 & 0.622 $\pm$ 0.242 & \textbf{0.914 $\pm$ 0.121} & 0.891 $\pm$ 0.143* \\
        Rouge-2 (High) & 0.014 $\pm$ 0.044 & 0.031 $\pm$ 0.084 & 0.132 $\pm$ 0.162 & 0.172 $\pm$ 0.183 & \textbf{0.615 $\pm$ 0.268} & 0.572 $\pm$ 0.272* \\
        \midrule
        Rouge-L (Low) & 0.608 $\pm$ 0.161 & 0.826 $\pm$ 0.190 & 0.927 $\pm$ 0.092 & 0.953 $\pm$ 0.072 & \textbf{0.989 $\pm$ 0.031} & 0.979 $\pm$ 0.042* \\
        Rouge-L (Mid) & 0.291 $\pm$ 0.144 & 0.458 $\pm$ 0.241 & 0.673 $\pm$ 0.197 & 0.765 $\pm$ 0.172 & \textbf{0.952 $\pm$ 0.070} & 0.939 $\pm$ 0.085* \\
        Rouge-L (High) & 0.111 $\pm$ 0.102 & 0.130 $\pm$ 0.135 & 0.313 $\pm$ 0.186 & 0.364 $\pm$ 0.190 & \textbf{0.749 $\pm$ 0.200} & 0.720 $\pm$ 0.210* \\
        \bottomrule
    \end{tabular}
    \vspace{0.2em}
    \caption{Performance using metrics -- BLEU, METEOR, ROUGE-1, ROUGE-2, and ROUGE-L -- for different models at varying noise factors on the Zoom dataset. Bolded numbers indicate the best performance (mean) in each row, while an asterisk (*) shows the second-best performance (mean) within the same metric.}
    \label{tab:metrics_noise_levels_zoom}
\end{table*}
From Table~\ref{tab:metrics_noise_levels_phone} and \ref{tab:metrics_noise_levels_zoom}, it is evident that B-CoAtNet performs poorly in the presence of noise. Without error mitigation techniques, relying solely on a classification model is inadequate.  
This outcome aligns with our expectations and underscores the need for LLMs to effectively mitigate errors.
Here, we show that incorporating LLMs for error detection and correction significantly improves performance. 
Within the Llama family, performance improves with larger model sizes. Among all tested LLMs, GPT-4o achieves the best results, particularly excelling under high noise conditions. 
\par 
The results indicate that LLMs are highly effective in detecting and correcting errors caused by noise in ASCA applications. GPT-4o, in particular, emerged as the most robust model, maintaining high performance across all metrics and noise levels. These findings underscore the potential of LLMs to enhance real-world applications where environmental noise poses a significant challenge.
\par 
Although GPT-4o achieves the best performance across all metrics, its major drawback is the high number of parameters and the long inference time it requires. GPT-4o has 200$\times$, 67$\times$, 25$\times$ more parameters compared to Llama-3.2-1B, Llama-3.2-3B, and Llama-3.2-8B, respectively. In the next section, we demonstrate that comparable accuracy to GPT-4o can be achieved by fine-tuning a lightweight Llama model.

\subsection{RQ3: Mitigating errors with Fine-tuned LLMs}
\label{sec:ft_results}
\begin{figure*}[t]
    \centering

    \includegraphics[width=0.20\textwidth]{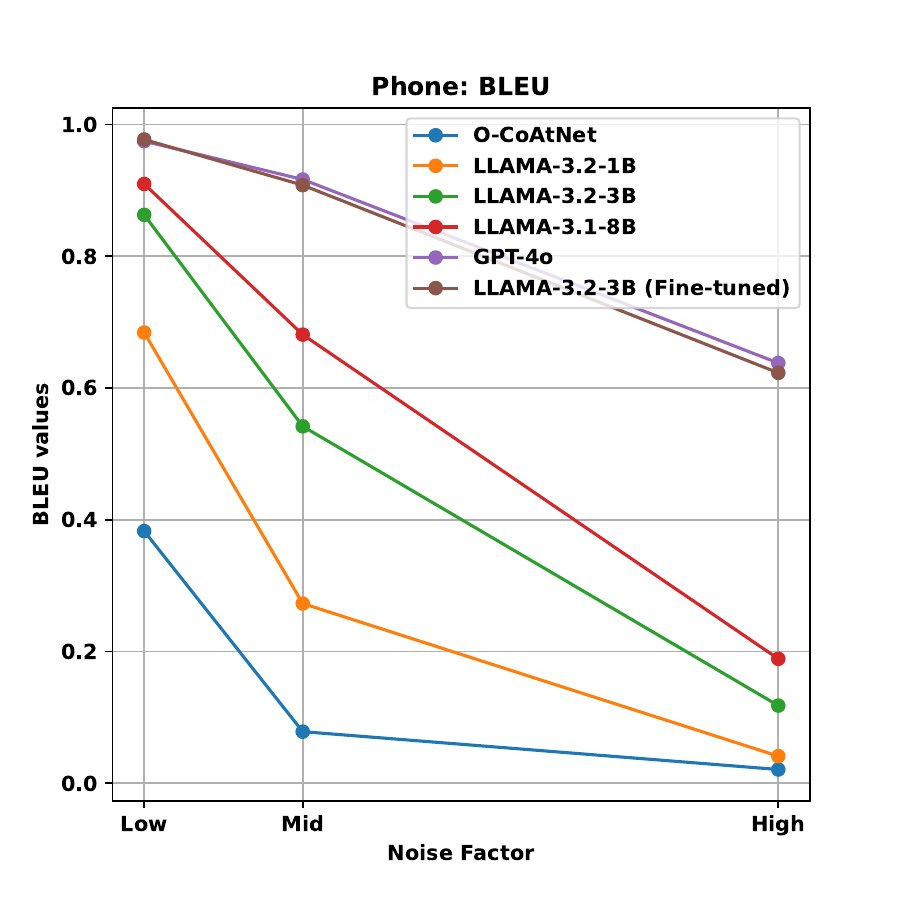}\hfill
    \includegraphics[width=0.20\textwidth]{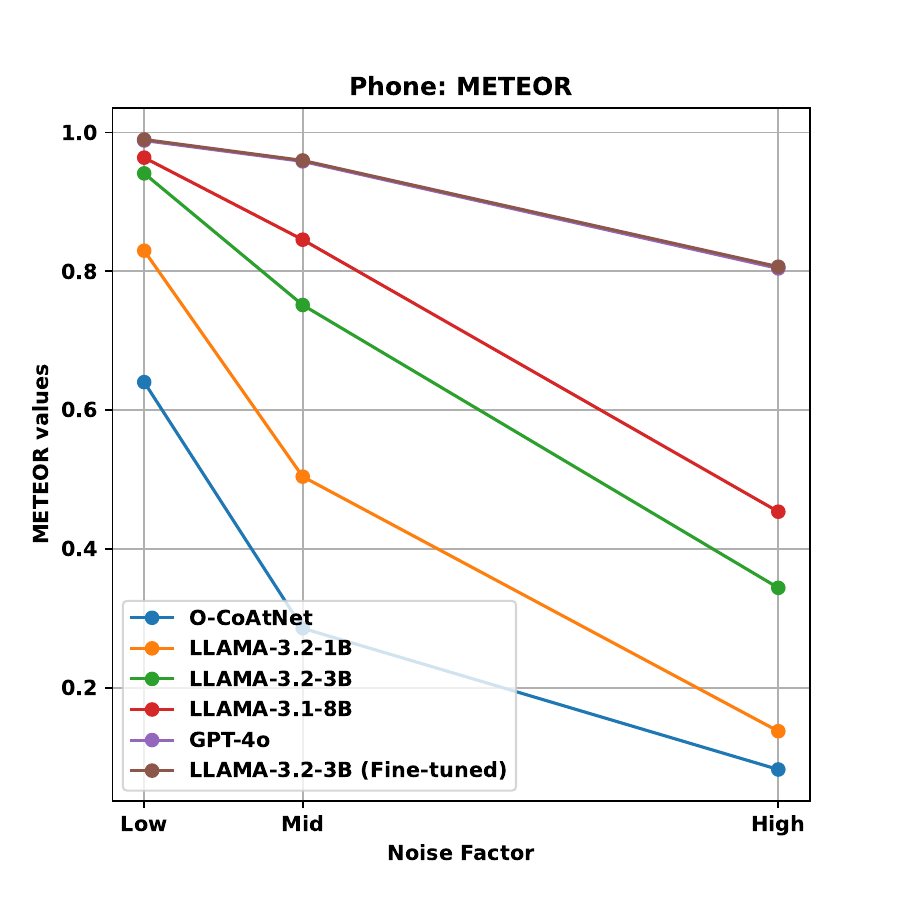}\hfill
    \includegraphics[width=0.20\textwidth]{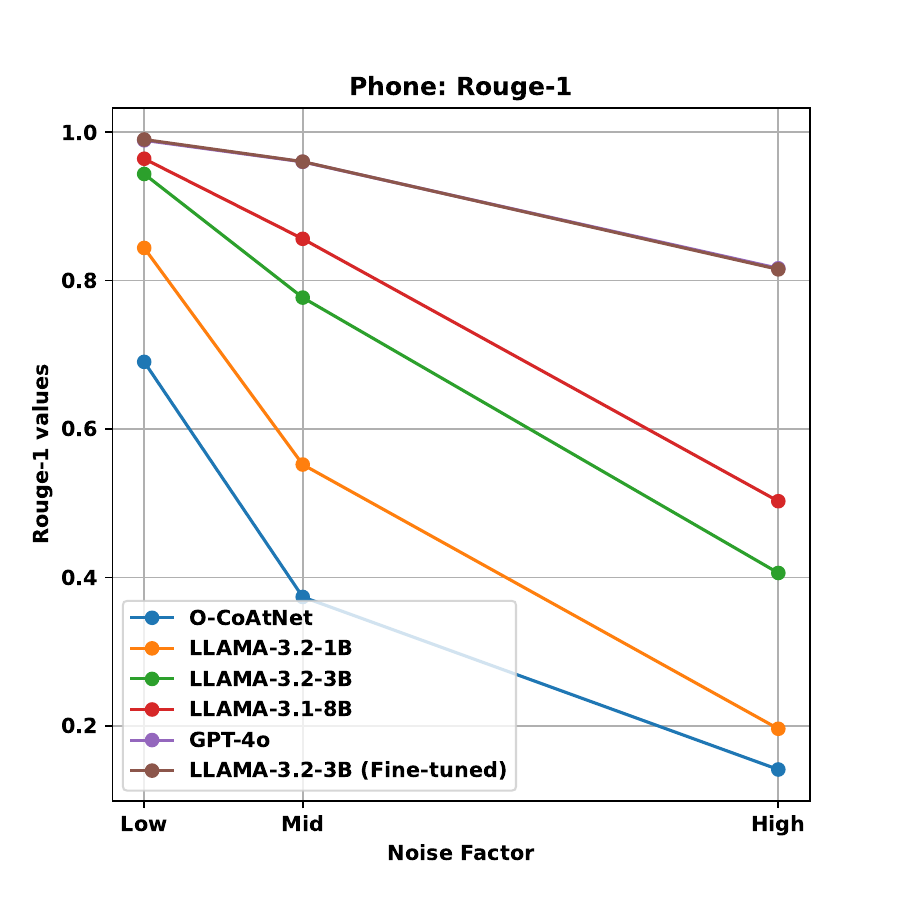}\hfill
    \includegraphics[width=0.20\textwidth]{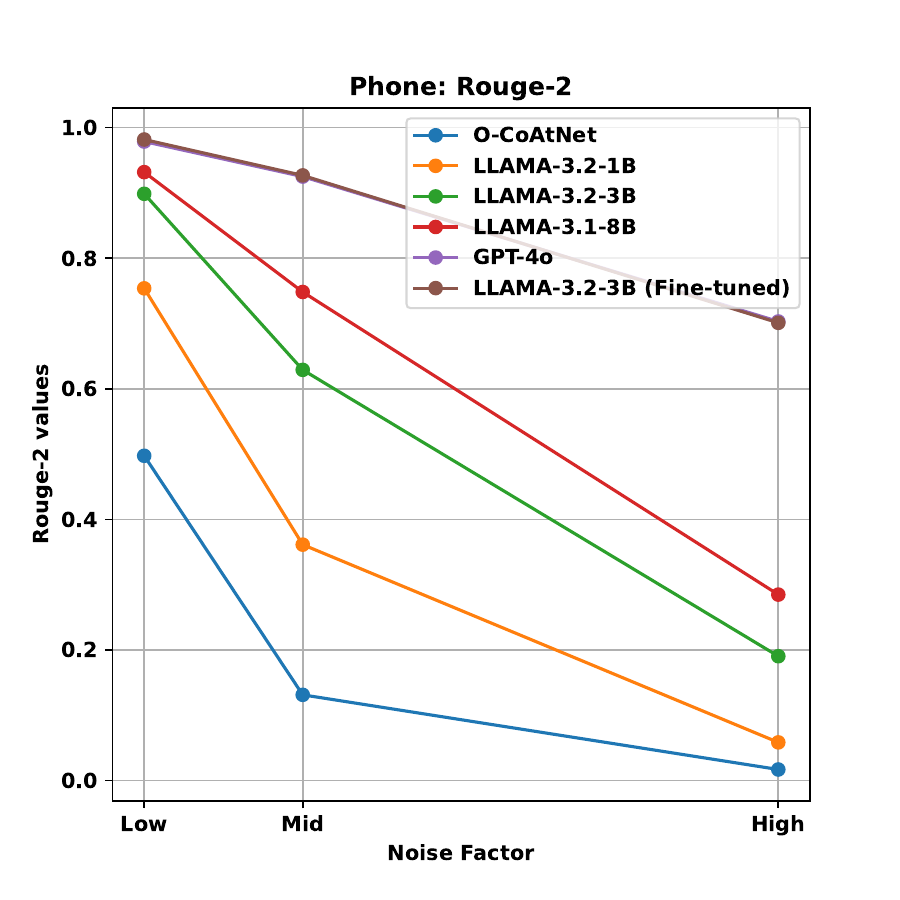}\hfill
    \includegraphics[width=0.20\textwidth]{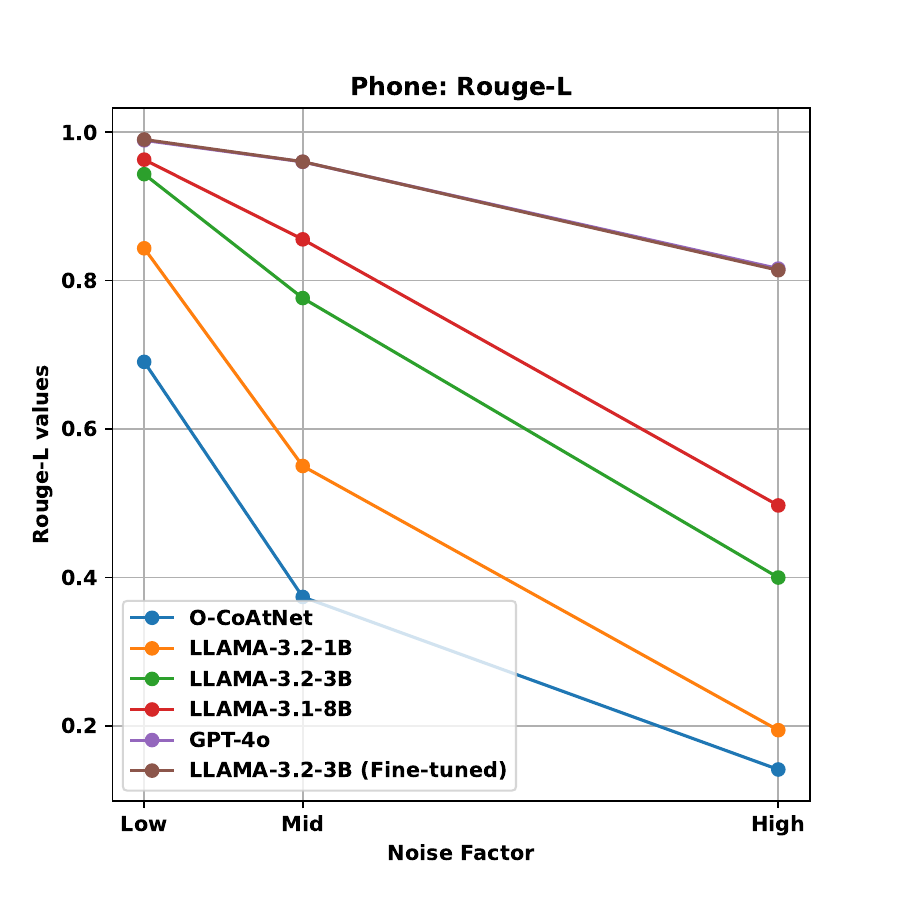}\hfill

    \includegraphics[width=0.20\textwidth]{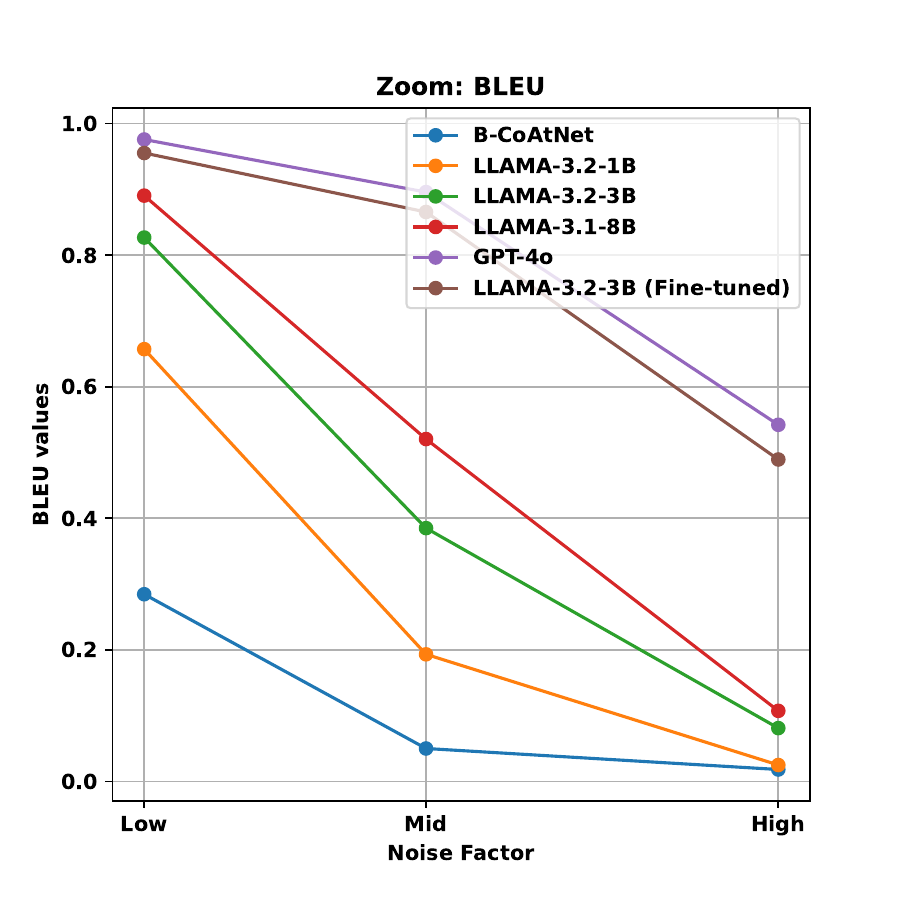}\hfill
    \includegraphics[width=0.20\textwidth]{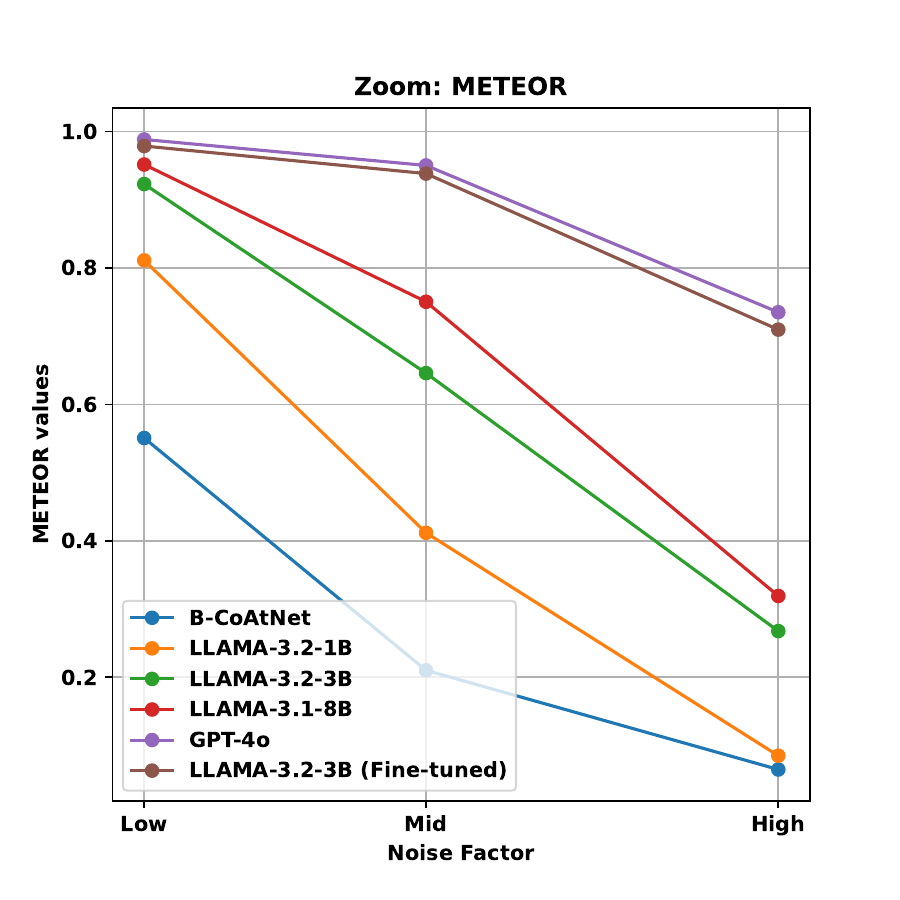}\hfill
    \includegraphics[width=0.20\textwidth]{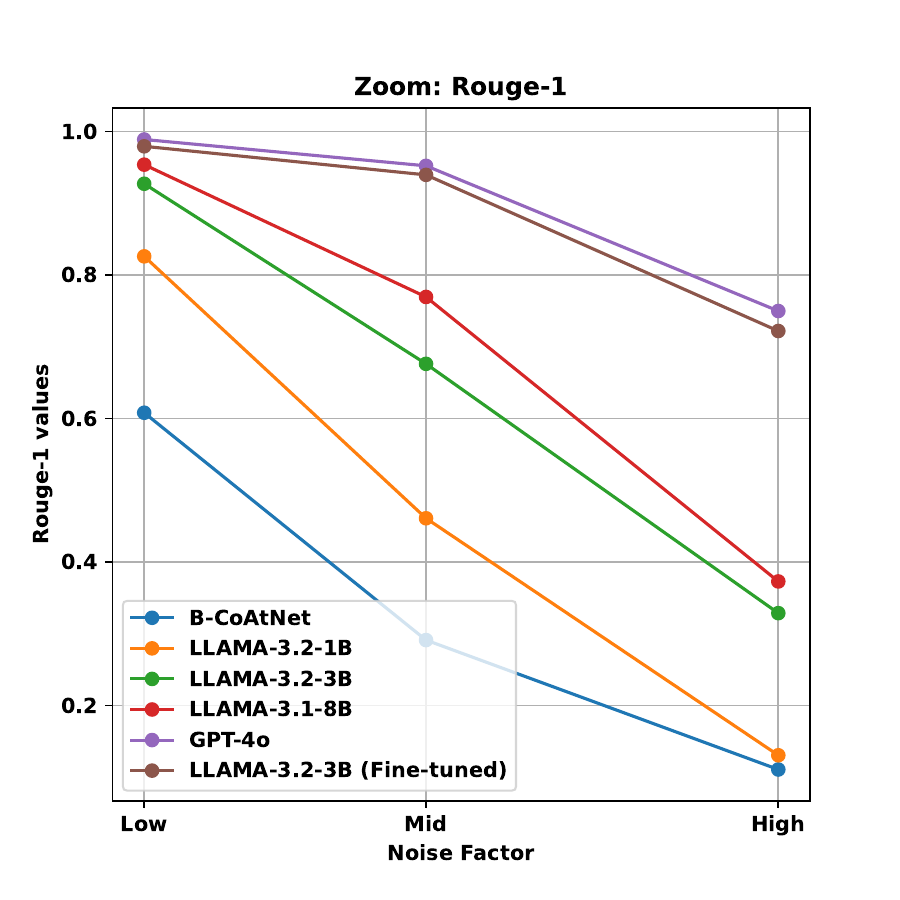}\hfill
    \includegraphics[width=0.20\textwidth]{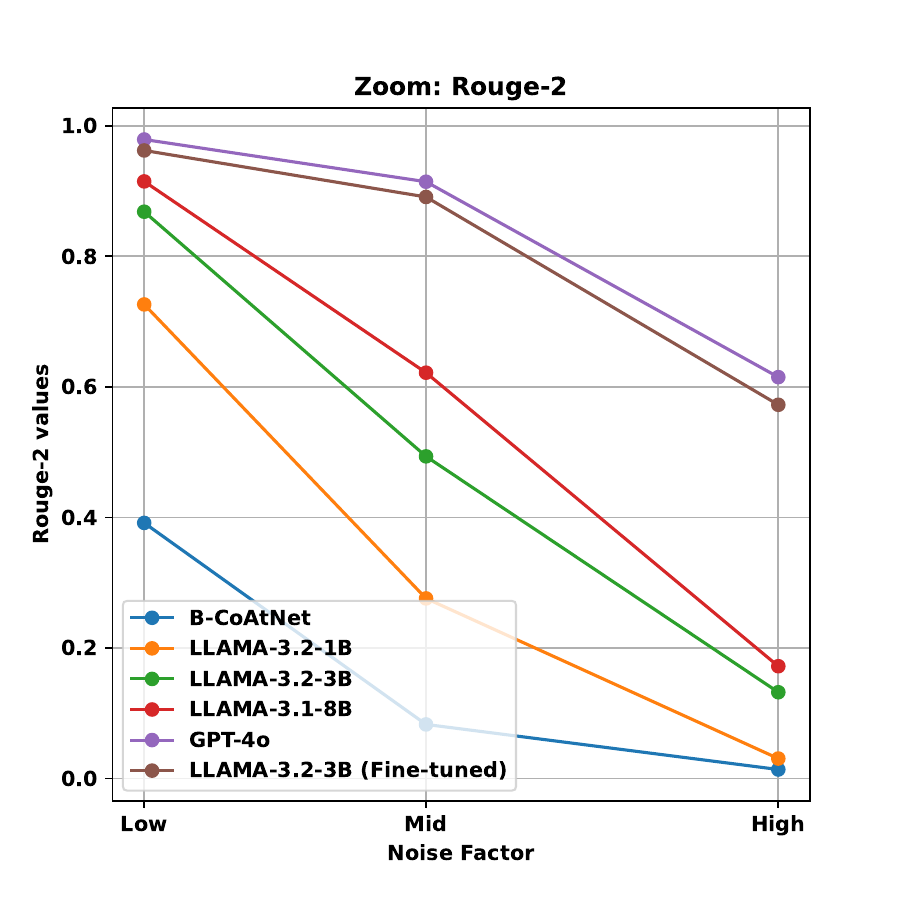}\hfill
    \includegraphics[width=0.20\textwidth]{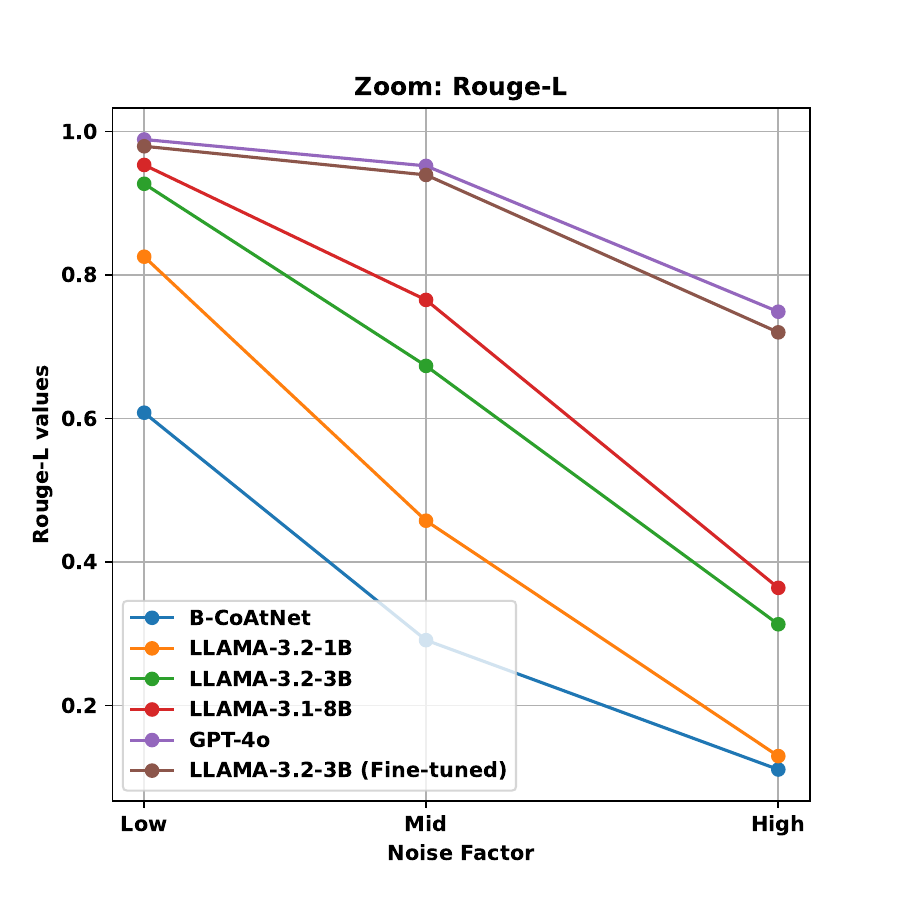}

    \caption{Performance using metrics -- BLEU, METEOR, ROUGE-1, ROUGE-2, and ROUGE-L -- for different models including the fine-tuned Llama-3.2-3B model at varying noise factors on the Phone and Zoom datasets. For clarity, only the mean is displayed in this graph; the standard deviation is omitted.}
    \label{fig:rotated}
\end{figure*}

Utilizing the same datasets as in the previous subsection, we fine-tuned the weights of Llama-3.2-3B and compared its performance with GPT-4o.
Similar to the previous analysis, we evaluated the performance using multiple metrics. See the rightmost column in Table~\ref{tab:metrics_noise_levels_phone} and \ref{tab:metrics_noise_levels_zoom}. Additionally, the results are visualized in Fig.~\ref{fig:rotated}. 
Both tables and graphs clearly demonstrate the fine-tuned model's strong ability to correct mistakes compared to other models. For the Phone dataset, the fine-tuned Llama outperforms GPT-4o when noise levels are moderate (Low and Mid). In contrast, for the Zoom dataset, GPT-4o consistently delivers superior performance under all conditions. Nonetheless, the fine-tuned model consistently ranks as the second-best performer.
\par
The fine-tuned model exhibits consistent performance improvements across all evaluation metrics. In comparison to the original Llama model without fine-tuning, the fine-tuned 3B model outperforms it by a significant margin. For instance, when performance gain is measured as the average percentage increase in BLEU accuracy across various noise levels, the fine-tuned 3B model shows nearly a 170\% improvement over the original 3B model for the Phone dataset. Moreover, it outperforms the larger 8B model by 90\%. A similar trend is observed on the Zoom dataset, with an 215\% improvement over the original 3B model and a 145\% increase compared to the 8B model. These results clearly demonstrate the substantial benefits of fine-tuning, enabling a smaller model to outperform even larger variants.
\par
The fine-tuning is especially effective at higher noise levels. For example, with the noise intensity set to High, the fine-tuning increased the BLEU accuracy by 430\% and 500\% on the phone and Zoom datasets, respectively, compared to the original 3B model 
Its effect is not as evident at Low noise intensity as they are already close to 1. 
Similar levels of improvement are observed in other metrics as well. 
For instance, in Fig.~\ref{fig:rotated}, the fine-tuned 3B model (in brown) performs considerably better than all the other Llama models (1B, 3B, 8B in orange, green, and red, respectively) under all evaluation metrics at higher noise settings. 
\par
Despite having only about 1.5\% of the parameter count of the much larger GPT-4o, the fine-tuned model delivers comparable performance, achieving at least 90\% of GPT-4o’s scores across all evaluation metrics.
Our fine-tuning aligns the small Llama model with the large GPT-4o model in terms of noise tolerance.
This indicates that a lightweight model like Llama-3.2-3B is sufficient to complete the error correction task.
In conclusion, we showed that the fine-tuning approach has proven to be highly effective, transforming the pipeline’s output from partially correct sentences to highly accurate predictions of the ground truth sentences.
\par

%% file: Sections/disc.tex
\section{Discussion}
\label{sec:discussion}

\paragraph{Our Contributions.}
Our results highlight a major weakness in prior ASCA approaches—their vulnerability to noise. While previous methods achieved high accuracy in noise-free conditions, their performance deteriorates in real-world settings. We address this gap by leveraging transformer-based architectures (VTs and LLMs) for robust classification and noise correction. Our findings underscore the need for ASCA research to prioritize noise resilience and adopt transformer-driven methodologies.

\paragraph{Implications of Findings.}
Our findings challenge the assumption that passphrases provide both security and usability against LLM-assisted ASCA attacks. While prior research shows passphrases are easier to remember and more secure than simple passwords \cite{keith2009behavioral}, LLMs can reconstruct them using linguistic patterns. However, they struggle with entirely random inputs, which are impractical for users. As transformer-based ASCAs grow more effective, both passwords and passphrases become unreliable. This reinforces the need to shift towards alternative authentication methods like hardware-based security, Multi-Factor Authentication (MFA), or behavioral biometrics, which are more resistant to side-channel inference.

\paragraph{This Work's Limitations.}
Our study has limitations, particularly in dataset size and scope. It includes only 25 samples per keystroke, limited to alphanumeric inputs, excluding essential keys like space, backspace, and enter. Recordings were restricted to a MacBook Pro keyboard, limiting generalizability across devices and typing styles. Additionally, we used synthetic Gaussian noise, whereas real-world environments involve more complex noise sources. While we mitigated these factors by varying noise levels and using standard keyboards, our key contribution lies in demonstrating the feasibility of transformer-based contextual error correction, paving the way for future noise-resilient ASCA research.

\paragraph{This Fields's Limitations.}
ASCA research lacks large, diverse, and public datasets, limiting progress and reproducibility. Unlike speech recognition or NLP, it relies on small, non-standardized datasets, making benchmarking difficult. The absence of varied keystroke-acoustic data hinders robust model development. To advance the field, community-driven dataset collection is crucial. Without standardized, large-scale datasets, progress will remain fragmented, and security risks harder to assess. We urge collaborative efforts to establish open ASCA benchmarks, similar to those in speech and image recognition, to drive systematic advancements.

\paragraph{Our Future Work.}
Given these limitations, future research has significant potential. Expanded datasets should include additional keyboard keys (function, space, backspace, arrows, special characters), diverse device models (desktop and soft keyboards on smartphones/tablets), and varied typing habits to improve generalization. Critically, experiments should incorporate authentic ambient noise (e.g., coffee shop chatter, street noise, music, office sounds) instead of synthetic models alone. While our study introduced realism by adjusting noise levels, future work should explicitly simulate diverse acoustic scenarios and thoroughly evaluate performance. Additionally, exploring real-time error correction with lightweight, fine-tuned LLMs can enhance robust side-channel methods on constrained edge devices, requiring efficient inference, low latency, and improved usability. We encourage prioritizing these directions to advance practical ASCA solutions.

\paragraph{Field's Future Work.}
Future ASCA research will extend transformer-based methods beyond acoustics into visual side-channels (screen reflections, typing videos) and electromagnetic emanations. As LLMs can reconstruct passphrases despite noise, password-based authentication may become less viable, prompting shifts toward multi-factor authentication, behavioral biometrics, or continuous authentication. Addressing the lack of large, standardized public datasets covering diverse keyboards, typing styles, and real-world noise will be critical for reproducibility. Finally, efficient transformer optimizations (LoRA, QLoRA, distillation) could enable real-time, on-device ASCA attacks, increasing security risks and driving demand for new cybersecurity countermeasures.


%% file: Sections/rel.tex
\section{Related Works}
\label{sec:related}

\noindent
\textbf{Acoustic Side-Channel Attacks (ASCAs)} exploit sound emanations from electronic devices, particularly keyboards, to infer sensitive information like passwords and PINs. With the ubiquity of microphones, these attacks have evolved over decades, addressing prior limitations while presenting new challenges and research opportunities. This section reviews their progression, key challenges, and areas for improvement.

\textbf{Traditional statistical ASCA approaches} relied on signal processing and acoustic feature analysis, using methods like cosine similarity and cross-correlation. Cosine similarity effectively modeled inter-keystroke timings to infer numeric inputs such as PIN codes~\cite{liu2019human}, while cross-correlation highlighted acoustic signal consistency across similar keystrokes~\cite{panda2020behavioral}. Techniques such as Time Difference of Arrival (TDoA) improved accuracy by leveraging geometric keyboard positions~\cite{de2019differential, zhu2014context}. Hidden Markov Models (HMMs), enhanced by timing and language-based post-processing (e.g., n-grams and spelling correction), further reconstructed text from keystroke acoustics~\cite{foo2010timing, zhuang2009keyboard}. Nevertheless, these methods lacked robustness, performed well primarily on limited numeric or predictable text, and struggled significantly with full alphanumeric recovery under realistic noise and typing variations~\cite{taheritajar2024survey}.

\textbf{Machine learning methods} enhanced ASCA robustness by using advanced models like clustering, Support Vector Machines (SVMs), and Hidden Markov Models (HMMs). K-means clustering combined with Mel-Frequency Cepstrum Coefficients (MFCCs) effectively grouped keystrokes into distinct acoustic classes~\cite{wit2014all}. Systematic use of HMMs~\cite{zhuang2009keyboard}, SVMs~\cite{wang2016accurate}, logistic regression, and random forests~\cite{anand2018keyboard} further improved accuracy in controlled environments. Despite these advances, machine learning methods remained highly sensitive to typing styles, keyboard variations, and acoustic noise, with accuracy dropping by 30–50\% under realistic conditions~\cite{halevi2012closer, fang2018no, liu2015snooping}.

\textbf{Recent deep learning (DL) architectures} significantly outperformed traditional methods by extracting hierarchical features from acoustic spectrograms. CNNs, ConvMixer models, and CNN-RNN hybrids achieved over 90\% keystroke classification accuracy under clean conditions~\cite{harrison2023practical, akinbi2023password, giallanza2019keyboard, toreini2015acoustic}; for instance, Harrison et al. reached above 93\% accuracy using CoAtNet models~\cite{harrison2023practical}. However, these DL methods remain constrained by their need for extensive labeled data, high computational costs, and notably poor performance (often below 40\%) in realistic noisy environments~\cite{giallanza2019keyboard}. This fundamental lack of robustness arises from purely local classification approaches, failing to leverage global contextual semantics, thus limiting their practical effectiveness in real-world ASCA scenarios.

\textbf{Transformer architectures} recently emerged as powerful methods capable of modeling global relationships and handling long-range dependencies, significantly improving noise robustness. Although transformers have demonstrated strong results in computer vision (Vision Transformers, VTs) and natural language processing (Large Language Models, LLMs), their application to acoustic side-channel attacks remains unexplored. VTs, through self-attention mechanisms, can identify global acoustic patterns, potentially addressing local noise challenges, while LLMs' proven contextual reasoning abilities~\cite{dosovitskiy2020image, brown2020language} could correct misclassifications in noisy keystroke scenarios. Nevertheless, transformer implementation in ASCA faces practical challenges, primarily computational costs, though recent lightweight fine-tuning methods like LoRA and QLoRA significantly alleviate these concerns~\cite{hu2021lora, dettmers2024qlora}. Yet, the effectiveness of these lightweight methods specifically for acoustic keystroke error correction has not been validated.

\textbf{In summary}, current ASCA methods are limited by their reliance on local acoustic classification, greatly reducing accuracy in noisy environments. Transformers, capable of modeling global dependencies and context, represent a promising yet unexplored solution to these limitations. Recent lightweight fine-tuning advancements (LoRA, QLoRA) further suggest practical feasibility. Here, we explicitly evaluate the hypothesis that integrating vision transformers and LLM-based contextual correction addresses noise robustness without sacrificing computational efficiency, conclusively determining their value in practical ASCA scenarios.


%% file: Sections/conc.tex
\section{Conclusion}
\label{sec:conclusion}
In this paper, we investigated the limitations of existing Acoustic Side-Channel Attack (ASCA) methods in controlled noisy noisy scenarios and demonstrated the potential of transformer architectures to address these challenges. By fine-tuning a CoAtNet model and introducing Vision Transformers (VTs), we established a new state-of-the-art benchmark in keystroke classification accuracy, surpassing traditional CNNs in clean acoustic conditions. Crucially, we showed that Large Language Models (LLMs) significantly mitigate errors caused by realistic noise, improving the reliability and practicality of ASCAs.
Furthermore, we found that lightweight LLMs fine-tuned with efficient techniques such as LoRA and QLoRA can achieve correction performance comparable to larger models, making them suitable even for resource-constrained attacks. Our results underline the importance of rigorously evaluating noise robustness in future ASCA research, highlighting transformer-based context-aware methods as essential for effective acoustic side-channel text recovery.
Finally, to enhance transparency and reproducibility, we will release our trained models, fine-tuning configurations, experimental pipeline code, and augmented acoustic-keyboard datasets with realistic noise scenarios as open-source resources. This aims to facilitate future research, encouraging the security community and practitioners to build on these findings toward comprehensive, noise-resilient ASCA solutions.

\noindent \textbf{Reproducibility.} Our source code is available at \url{https://github.com/seyyedaliayati/EchoCrypt}, and our fine-tuned model weights are available at \url{https://huggingface.co/seyyedaliayati/zoom_model} and \url{https://huggingface.co/seyyedaliayati/phone_model}.

\noindent \textbf{Acknowledgments.} We would like to thank the NSF for the support via the CNS 2327427 grant.

%% file: main.bbl
\begin{thebibliography}{10}

\bibitem{achiam2023gpt}
{\sc Achiam, J., Adler, S., Agarwal, S., Ahmad, L., Akkaya, I., Aleman, F.~L., Almeida, D., Altenschmidt, J., Altman, S., Anadkat, S., et~al.}
\newblock Gpt-4 technical report.
\newblock {\em arXiv preprint arXiv:2303.08774\/} (2023).

\bibitem{akinbi2023password}
{\sc Akinbi, A., Deniz, E., Ismael, A.~M., Rashid, Z.~N., and Sengur, A.}
\newblock Password-sniffing acoustic keylogger using machine learning.
\newblock {\em Available at SSRN 4431909\/} (2023).

\bibitem{anand2018keyboard}
{\sc Anand, S.~A., and Saxena, N.}
\newblock Keyboard emanations in remote voice calls: Password leakage and noise (less) masking defenses.
\newblock In {\em Proceedings of the Eighth ACM Conference on Data and Application Security and Privacy\/} (2018), pp.~103--110.

\bibitem{asonov2004keyboard}
{\sc Asonov, D., and Agrawal, R.}
\newblock Keyboard acoustic emanations.
\newblock In {\em IEEE Symposium on Security and Privacy, 2004. Proceedings. 2004\/} (2004), IEEE, pp.~3--11.

\bibitem{ayman2024englishtense}
{\sc Ayman, U., Rahman, M.~H., and Islam, M.~S.}
\newblock {EnglishTense: A large scale English texts dataset categorized into three categories: Past, Present, Future tenses}, 2024.

\bibitem{backes2010acoustic}
{\sc Backes, M., D{\"u}rmuth, M., Gerling, S., Pinkal, M., Sporleder, C., et~al.}
\newblock Acoustic $\{$Side-Channel$\}$ attacks on printers.
\newblock In {\em 19th USENIX Security Symposium (USENIX Security 10)\/} (2010).

\bibitem{bao2021beit}
{\sc Bao, H., Dong, L., Piao, S., and Wei, F.}
\newblock Beit: Bert pre-training of image transformers.
\newblock {\em arXiv preprint arXiv:2106.08254\/} (2021).

\bibitem{berger2006dictionary}
{\sc Berger, Y., Wool, A., and Yeredor, A.}
\newblock Dictionary attacks using keyboard acoustic emanations.
\newblock In {\em Proceedings of the 13th ACM conference on Computer and communications security\/} (2006), pp.~245--254.

\bibitem{brown2020language}
{\sc Brown, T., Mann, B., Ryder, N., Subbiah, M., Kaplan, J.~D., Dhariwal, P., Neelakantan, A., Shyam, P., Sastry, G., Askell, A., et~al.}
\newblock Language models are few-shot learners.
\newblock {\em Advances in neural information processing systems 33\/} (2020), 1877--1901.

\bibitem{carrara2015acoustic}
{\sc Carrara, B., and Adams, C.}
\newblock On acoustic covert channels between air-gapped systems.
\newblock In {\em Foundations and Practice of Security: 7th International Symposium, FPS 2014, Montreal, QC, Canada, November 3-5, 2014. Revised Selected Papers 7\/} (2015), Springer, pp.~3--16.

\bibitem{compagno2017don}
{\sc Compagno, A., Conti, M., Lain, D., and Tsudik, G.}
\newblock Don't skype \& type! acoustic eavesdropping in voice-over-ip.
\newblock In {\em Proceedings of the 2017 ACM on Asia Conference on Computer and Communications Security\/} (2017), pp.~703--715.

\bibitem{dai2021coatnet}
{\sc Dai, Z., Liu, H., Le, Q.~V., and Tan, M.}
\newblock Coatnet: Marrying convolution and attention for all data sizes.
\newblock {\em Advances in neural information processing systems 34\/} (2021), 3965--3977.

\bibitem{unsloth}
{\sc Daniel~Han, M.~H., and team, U.}
\newblock Unsloth, 2023.

\bibitem{de2019differential}
{\sc de~Souza~Faria, G., and Kim, H.~Y.}
\newblock Differential audio analysis: a new side-channel attack on pin pads.
\newblock {\em International Journal of Information Security 18\/} (2019), 73--84.

\bibitem{5206848}
{\sc Deng, J., Dong, W., Socher, R., Li, L.-J., Li, K., and Fei-Fei, L.}
\newblock Imagenet: A large-scale hierarchical image database.
\newblock In {\em 2009 IEEE Conference on Computer Vision and Pattern Recognition\/} (2009), pp.~248--255.

\bibitem{dettmers2023qloraefficientfinetuningquantized}
{\sc Dettmers, T., Pagnoni, A., Holtzman, A., and Zettlemoyer, L.}
\newblock Qlora: Efficient finetuning of quantized llms, 2023.

\bibitem{dettmers2024qlora}
{\sc Dettmers, T., Pagnoni, A., Holtzman, A., and Zettlemoyer, L.}
\newblock Qlora: Efficient finetuning of quantized llms.
\newblock {\em Advances in Neural Information Processing Systems 36\/} (2024).

\bibitem{dosovitskiy2020image}
{\sc Dosovitskiy, A.}
\newblock An image is worth 16x16 words: Transformers for image recognition at scale.
\newblock {\em arXiv preprint arXiv:2010.11929\/} (2020).

\bibitem{esmaeilpour2020sound}
{\sc Esmaeilpour, M., Cardinal, P., and Koerich, A.~L.}
\newblock From sound representation to model robustness.
\newblock {\em arXiv preprint arXiv:2007.13703\/} (2020).

\bibitem{fang2018no}
{\sc Fang, S., Markwood, I., Liu, Y., Zhao, S., Lu, Z., and Zhu, H.}
\newblock No training hurdles: Fast training-agnostic attacks to infer your typing.
\newblock In {\em Proceedings of the 2018 ACM SIGSAC Conference on Computer and Communications Security\/} (2018), pp.~1747--1760.

\bibitem{foo2010timing}
{\sc Foo~Kune, D., and Kim, Y.}
\newblock Timing attacks on pin input devices.
\newblock In {\em Proceedings of the 17th ACM conference on Computer and communications security\/} (2010), pp.~678--680.

\bibitem{giallanza2019keyboard}
{\sc Giallanza, T., Siems, T., Smith, E., Gabrielsen, E., Johnson, I., Thornton, M.~A., and Larson, E.~C.}
\newblock Keyboard snooping from mobile phone arrays with mixed convolutional and recurrent neural networks.
\newblock {\em Proceedings of the ACM on Interactive, Mobile, Wearable and Ubiquitous Technologies 3}, 2 (2019), 1--22.

\bibitem{halevi2012closer}
{\sc Halevi, T., and Saxena, N.}
\newblock A closer look at keyboard acoustic emanations: random passwords, typing styles and decoding techniques.
\newblock In {\em Proceedings of the 7th ACM Symposium on Information, Computer and Communications Security\/} (2012), pp.~89--90.

\bibitem{hannun2014deep}
{\sc Hannun, A., Case, C., Casper, J., Catanzaro, B., Diamos, G., Elsen, E., Prenger, R., Satheesh, S., Sengupta, S., Coates, A., et~al.}
\newblock Deep speech: Scaling up end-to-end speech recognition.
\newblock {\em arXiv preprint arXiv:1412.5567\/} (2014).

\bibitem{harrison2023practical}
{\sc Harrison, J., Toreini, E., and Mehrnezhad, M.}
\newblock A practical deep learning-based acoustic side channel attack on keyboards.
\newblock In {\em 2023 IEEE European Symposium on Security and Privacy Workshops (EuroS\&PW)\/} (2023), IEEE, pp.~270--280.

\bibitem{hu2021lora}
{\sc Hu, E.~J., Shen, Y., Wallis, P., Allen-Zhu, Z., Li, Y., Wang, S., Wang, L., and Chen, W.}
\newblock Lora: Low-rank adaptation of large language models.
\newblock {\em arXiv preprint arXiv:2106.09685\/} (2021).

\bibitem{hurst2024gpt}
{\sc Hurst, A., Lerer, A., Goucher, A.~P., Perelman, A., Ramesh, A., Clark, A., Ostrow, A., Welihinda, A., Hayes, A., Radford, A., et~al.}
\newblock Gpt-4o system card.
\newblock {\em arXiv preprint arXiv:2410.21276\/} (2024).

\bibitem{datasetgithublink}
{\sc JBFH-Dev}.
\newblock Keystroke-datasets.
\newblock \url{https://github.com/JBFH-Dev/Keystroke-Datasets}, 2023.

\bibitem{keith2009behavioral}
{\sc Keith, M., Shao, B., and Steinbart, P.}
\newblock A behavioral analysis of passphrase design and effectiveness.
\newblock {\em Journal of the Association for Information Systems 10}, 2 (2009), 2.

\bibitem{kingma2014adam}
{\sc Kingma, D.~P.}
\newblock Adam: A method for stochastic optimization.
\newblock {\em arXiv preprint arXiv:1412.6980\/} (2014).

\bibitem{liu2015snooping}
{\sc Liu, J., Wang, Y., Kar, G., Chen, Y., Yang, J., and Gruteser, M.}
\newblock Snooping keystrokes with mm-level audio ranging on a single phone.
\newblock In {\em Proceedings of the 21st Annual International Conference on Mobile Computing and Networking\/} (2015), pp.~142--154.

\bibitem{liu2019human}
{\sc Liu, X., Li, Y., Deng, R.~H., Chang, B., and Li, S.}
\newblock When human cognitive modeling meets pins: User-independent inter-keystroke timing attacks.
\newblock {\em Computers \& Security 80\/} (2019), 90--107.

\bibitem{liu2021swin}
{\sc Liu, Z., Lin, Y., Cao, Y., Hu, H., Wei, Y., Zhang, Z., Lin, S., and Guo, B.}
\newblock Swin transformer: Hierarchical vision transformer using shifted windows.
\newblock In {\em Proceedings of the IEEE/CVF international conference on computer vision\/} (2021), pp.~10012--10022.

\bibitem{loshchilov2017decoupled}
{\sc Loshchilov, I.}
\newblock Decoupled weight decay regularization.
\newblock {\em arXiv preprint arXiv:1711.05101\/} (2017).

\bibitem{mehrnezhad2018stealing}
{\sc Mehrnezhad, M., Toreini, E., Shahandashti, S.~F., and Hao, F.}
\newblock Stealing pins via mobile sensors: actual risk versus user perception.
\newblock {\em International Journal of Information Security 17}, 3 (2018), 291--313.

\bibitem{ninan2024second}
{\sc Ninan, M., Nimmo, E., Reilly, S., Smith, C., Sun, W., Wang, B., and Emmert, J.~M.}
\newblock A second look at the portability of deep learning side-channel attacks over em traces.
\newblock In {\em Proceedings of the 27th International Symposium on Research in Attacks, Intrusions and Defenses\/} (2024), pp.~630--643.

\bibitem{panda2020behavioral}
{\sc Panda, S., Liu, Y., Hancke, G.~P., and Qureshi, U.~M.}
\newblock Behavioral acoustic emanations: Attack and verification of pin entry using keypress sounds.
\newblock {\em Sensors 20}, 11 (2020), 3015.

\bibitem{radford2021learning}
{\sc Radford, A., Kim, J.~W., Hallacy, C., Ramesh, A., Goh, G., Agarwal, S., Sastry, G., Askell, A., Mishkin, P., Clark, J., et~al.}
\newblock Learning transferable visual models from natural language supervision.
\newblock In {\em International conference on machine learning\/} (2021), PMLR, pp.~8748--8763.

\bibitem{radford2023robust}
{\sc Radford, A., Kim, J.~W., Xu, T., Brockman, G., McLeavey, C., and Sutskever, I.}
\newblock Robust speech recognition via large-scale weak supervision.
\newblock In {\em International conference on machine learning\/} (2023), PMLR, pp.~28492--28518.

\bibitem{shamir2004acoustic}
{\sc Shamir, A., and Tromer, E.}
\newblock Acoustic cryptanalysis, 2004.
\newblock Accessed: 2025-03-08.

\bibitem{sun2017revisiting}
{\sc Sun, C., Shrivastava, A., Singh, S., and Gupta, A.}
\newblock Revisiting unreasonable effectiveness of data in deep learning era.
\newblock In {\em Proceedings of the IEEE international conference on computer vision\/} (2017), pp.~843--852.

\bibitem{taheritajar2024survey}
{\sc Taheritajar, A., Harris, Z.~M., and Rahaeimehr, R.}
\newblock A survey on acoustic side channel attacks on keyboards.
\newblock In {\em International Conference on Information and Communications Security\/} (2024), Springer, pp.~99--121.

\bibitem{toreini2015acoustic}
{\sc Toreini, E., Randell, B., and Hao, F.}
\newblock An acoustic side channel attack on enigma.
\newblock {\em School of Computing Science Technical Report Series\/} (2015).

\bibitem{touvron2021training}
{\sc Touvron, H., Cord, M., Douze, M., Massa, F., Sablayrolles, A., and J{\'e}gou, H.}
\newblock Training data-efficient image transformers \& distillation through attention.
\newblock In {\em International conference on machine learning\/} (2021), PMLR, pp.~10347--10357.

\bibitem{touvron2023llama}
{\sc Touvron, H., Lavril, T., Izacard, G., Martinet, X., Lachaux, M.-A., Lacroix, T., Rozi{\`e}re, B., Goyal, N., Hambro, E., Azhar, F., et~al.}
\newblock Llama: Open and efficient foundation language models.
\newblock {\em arXiv preprint arXiv:2302.13971\/} (2023).

\bibitem{wang2016accurate}
{\sc Wang, J., Ruby, R., Wang, L., and Wu, K.}
\newblock Accurate combined keystrokes detection using acoustic signals.
\newblock In {\em 2016 12th International Conference on Mobile Ad-Hoc and Sensor Networks (MSN)\/} (2016), IEEE, pp.~9--14.

\bibitem{wit2014all}
{\sc Wit, E., and Houtenbos, T.}
\newblock All your keystrokes are belong to us.
\newblock {\em Academia. edu\/} (2014).
\newblock Accessed 2025-03-08.

\bibitem{yang2023slnet}
{\sc Yang, Z., Zhang, Y., Qian, K., and Wu, C.}
\newblock $\{$SLNet$\}$: A spectrogram learning neural network for deep wireless sensing.
\newblock In {\em 20th USENIX Symposium on Networked Systems Design and Implementation (NSDI 23)\/} (2023), pp.~1221--1236.

\bibitem{zhu2014context}
{\sc Zhu, T., Ma, Q., Zhang, S., and Liu, Y.}
\newblock Context-free attacks using keyboard acoustic emanations.
\newblock In {\em Proceedings of the 2014 ACM SIGSAC conference on computer and communications security\/} (2014), pp.~453--464.

\bibitem{zhuang2009keyboard}
{\sc Zhuang, L., Zhou, F., and Tygar, J.~D.}
\newblock Keyboard acoustic emanations revisited.
\newblock {\em ACM Transactions on Information and System Security (TISSEC) 13}, 1 (2009), 1--26.

\end{thebibliography}
